\documentclass[aps,prb,reprint,showpacs,showkeys,groupedaddress]{revtex4-1}

\usepackage{multirow}
\usepackage{graphicx}
\usepackage{bm}
\usepackage{dcolumn}
\usepackage[hidelinks]{hyperref}
\usepackage{ulem}
\usepackage{natbib}
\begin{document}

\title{Finite-size effects in the quasi-one-dimensional quantum magnets Sr$_{2}$CuO$_{3}$, Sr$_{2}$Cu$_{0.99}M_{0.01}$O$_{3}$ ($M =$ Ni, Zn), and SrCuO$_{2}$}
\author{Koushik Karmakar}
\author{Surjeet Singh}
\email[]{surjeet.singh@iiserpune.ac.in}
\affiliation{Indian Institute of Science Education and Research,\\ Dr. Homi Bhabha Road, Pune, Maharashtra-411008, India}

\date{\today}

\begin{abstract}
We studied the finite-size effects on the magnetic behavior of the quasi-one-dimensional spin S = $\frac{1}{2}$ Heisenberg antiferromagnets Sr$_{2}$CuO$_{3}$, Sr$_{2}$Cu$_{0.99}M_{0.01}$O$_{3}$ ($M =$ Zn and Ni), and SrCuO$_{2}$. Magnetic susceptibility data were analyzed to estimate the concentration of chain breaks due to extrinsically doped defects and/or due to slight oxygen off-stoichiometry. We show that the susceptibility of Sr$_{2}$Cu$_{0.99}$Ni$_{0.01}$O$_{3}$ can be described by considering Ni$^{2+}$ as a scalar defect ($S_{eff}=0$) indicating that the Ni spin is screened. In Sr$_{2}$Cu$_{0.99}$Zn$_{0.01}$O$_{3}$ susceptibility analysis yields a defect concentration smaller than the nominal value which is in good qualitative agreement with crystal growth experiments. Influence of doping on the low-temperature long-range spin ordered state is studied. In the compound SrCuO$_{2}$, consisting of zigzag S = $\frac{1}{2}$ chains, the influence of spin frustration on the magnetic ordering and the defect concentration determined from the susceptibility data is discussed.
\end{abstract}

\maketitle
\section{Introduction}
One-dimensional (1D) spin S = $\frac{1}{2}$ Heisenberg antiferromagnetic (HAF) model is the cornerstone of our understanding of quantum magnets. While the exact eigenvalues and eigenstates of this model can in principle be obtained using the Bethe ansatz \cite{Bethe01}, calculations of thermodynamic quantities at nonzero temperatures has remained an active area of research. Probably, the first real attempt to calculate the susceptibility $\chi(T)$ and the specific heat $C_{p}(T)$ of the S = $\frac{1}{2}$ HAF chain was by Bonner and Fischer (BF) \cite{Bonner02}, who obtained numerical results for chains containing up to 11 spins subject to a periodic boundary condition and for temperatures $T \geq 0.4 J/k_{B}$ (where J is the intrachain nearest neighbor exchange and $k_{B}$ is the Boltzmann constant). These results revealed a broad susceptibility maximum near $T = 0.675 J/k_{B}$. At lower temperatures, the susceptibility decreases below the maximum value due to the gradual appearance of large antiferromagnetic domains characterized by domain-wall-like S = $\frac{1}{2}$ excitations called spinons. Due to quantum fluctuations, the domain-wall dynamics persist down to zero temperature. Therefore, the ground state of the S = $\frac{1}{2}$ HAF chain is characterized as a spin-liquid in which the spin correlations decay as a power law \cite{Balents03} and the zero-temperature susceptibility $\chi(0)$ has a finite value, which can be calculated exactly using the Bethe ansatz \cite{Griffiths04}. The $\chi(T)$ behavior at low temperatures, where the BF results are not accurate, was investigated by Eggert, Affleck, and Takahashi (EAT) using numerical Bethe ansatz ($T \geq 0.003 J/k_{B}$) and field-theory methods \cite{Eggert22}. The resulting $\chi(T)$ showed a surprisingly rich behavior with the existence of an inflection point near $T = 0.087 J/k_{B}$, below which the slope of $\chi(T)$ increases rapidly approaching infinity as T goes to zero. The field theory revealed a logarithmic leading order correction to the zero-temperature susceptibility $\chi(0)$. The accuracy of this result was further improved in the subsequent works by Lukyanov using field theory \cite{Lukyanov06}, and Kl\"umper and Johnston using the numerical Bethe ansatz \cite{Klumper05, Klumper33}. 

Experimental confirmation of this logarithmic behavior is limited, because in most quasi-1D systems either the intrachain interaction is too small, making the temperature range of interest difficult to access, or the system undergoes a dimensional crossover at low temperatures due to interchain interactions (J') (see, for example, Table 3 in ref. \citenum{Janson34}). Even in those rare cases where an excellent one-dimensionality and exceptionally large J can be combined into a single material, the inevitable presence of chain-breaks vitiate the ideal chain behavior at low temperatures. Due to chain breaks a chain is cut into finite-length segments. The staggered moment at the boundaries of these segments contributes a Curie-like susceptibility with an effective paramagnetic Curie constant that depends on temperature \cite{Fujimoto19}. It has been argued that in a nominally pure crystal additional paramagnetic impurities play a minor role in giving rise to the Curie tail \cite{Sirker09}. That the boundary effects are significant is also apparent from the fact that the measured susceptibility of powdered samples of quasi-1D chains can be significantly larger than that of the single crystals (for comparison see, for example, the susceptibility data published in Refs. \citenum{Ami07} and \citenum{Chattopadhyay35} on powder samples and that in Ref. \citenum{Motoyama08} on single crystals of the same quasi-1D compounds). Therefore, in analyzing the experimental data of a nominally pure spin chain compound or of one doped with a dilute level of “non-magnetic” impurities within the chains, it is important to take the boundary contribution into account for an accurate description of the experimental susceptibility. Conversely, knowledge of the boundary susceptibility allows for an accurate estimate of the defect concentration in a spin chain, which can otherwise be substantially underestimated\cite{Sirker09}. Since any real systemmust have boundaries due to defects (or chain breaks), it is pertinent to investigate and understand the effect of boundaries on the physical properties and to test the validity of the theoretical models that describe the boundary effects in spin chains.
   
In this paper we examined the finite-size effects in quasi one-dimensional Sr$_{2}$CuO$_{3}$ and SrCuO$_{2}$. The main objective of the present study is to carry out an independent verification of the field theory result by Sirker \textit{et al.} \cite{Sirker09}, which quantifies the effect of boundaries on the susceptibility of a spin chain. We also investigated the effective spin state of Ni$^{2+}$ ion doped in a spin $\frac{1}{2}$ chain. This question is important because the field theory result by Eggert and Affleck\cite{Eggert10} predicts that an isolated spin 1 impurity in a spin $\frac{1}{2}$ chain may essentially behaves like a scalar impurity (S$_{eff}$ = 0) due to the screening cloud formed by its antiferromagnetic coupling to its neighbors.

The plan of the paper is as follows: we first provide the experimental details, including a brief summary of the results of our crystal growth experiments which will be useful in the later part of the manuscript that deals with the data analysis and determination of defect concentration in the doped crystals. The susceptibility of pristine Sr$_{2}$CuO$_{3}$, containing intrinsic chain breaks arising due to oxygen off-stoichiometry, is analyzed first as it presents the simplest case due to its linear chain geometry and also allows for a comparison of our results with the previous study by Sirker \textit{et al.} \cite{Sirker09}. The procedure is then extended to analyze the susceptibility of pristine SrCuO$_{2}$ (zigzag chains) and the Zn and Ni doped Sr$_{2}$CuO$_{3}$ crystals. In the zigzag chains, the role of spin frustration on defect concentration determination is investigated. In the Zn doped Sr$_{2}$CuO$_{3}$ crystals, the defect concentration obtained is discussed in light of the crystal growth experiment results. The magnetization of the Ni-doped crystal is fitted by assuming effective Ni$^{2+}$ spin to be zero. The influence of Ni and Zn doping on the low-temperature phase transition in Sr$_{2}$CuO$_{3}$ is presented in the last section.

\section{Experimental Details}

The investigations were carried out on high-quality single crystals grown using a four-mirror infrared image furnace (Crystal System Corporation, Japan).   For the single crystal growth of Sr$_{2}$CuO$_{3}$ and its doped variants, the starting precursors used and their associated purities (indicated in the parenthesis) were SrCO$_{3}$ (99.99\%), CuO (99.995\%), ZnO (99.99\%), and NiO (99.99\%) and in the preparation of SrCuO$_{2}$ the precursors used were SrCO$_{3}$ (99.9\%) and CuO (99.99\%).  All the successfully grown crystals were characterized using optical microscopy under polarized light (Carl Zeiss), scanning electron microscopy (SEM) (Zeiss Ultra Plus), energy dispersive x-ray (EDX) analysis (Oxford), and x-ray powder diffraction (Bruker D8 Advance) obtained by grinding the crystal pieces. The crystals were oriented using a Laue camera equipped with microfocal x-ray source (Photonics Science). The dc magnetization measurements were carried out along the three principal axes of the grown crystals using a vibrating sample magnetometer (VSM) in a physical property measurement system (Quantum Design, USA). Prior to magnetic measurements the crystal specimens were annealed under Ar-flow for 72 h at 875$^{\circ}$C. 

\section{Crystal Growth and Structural Details}

\begin{figure}
\includegraphics[scale=0.5]{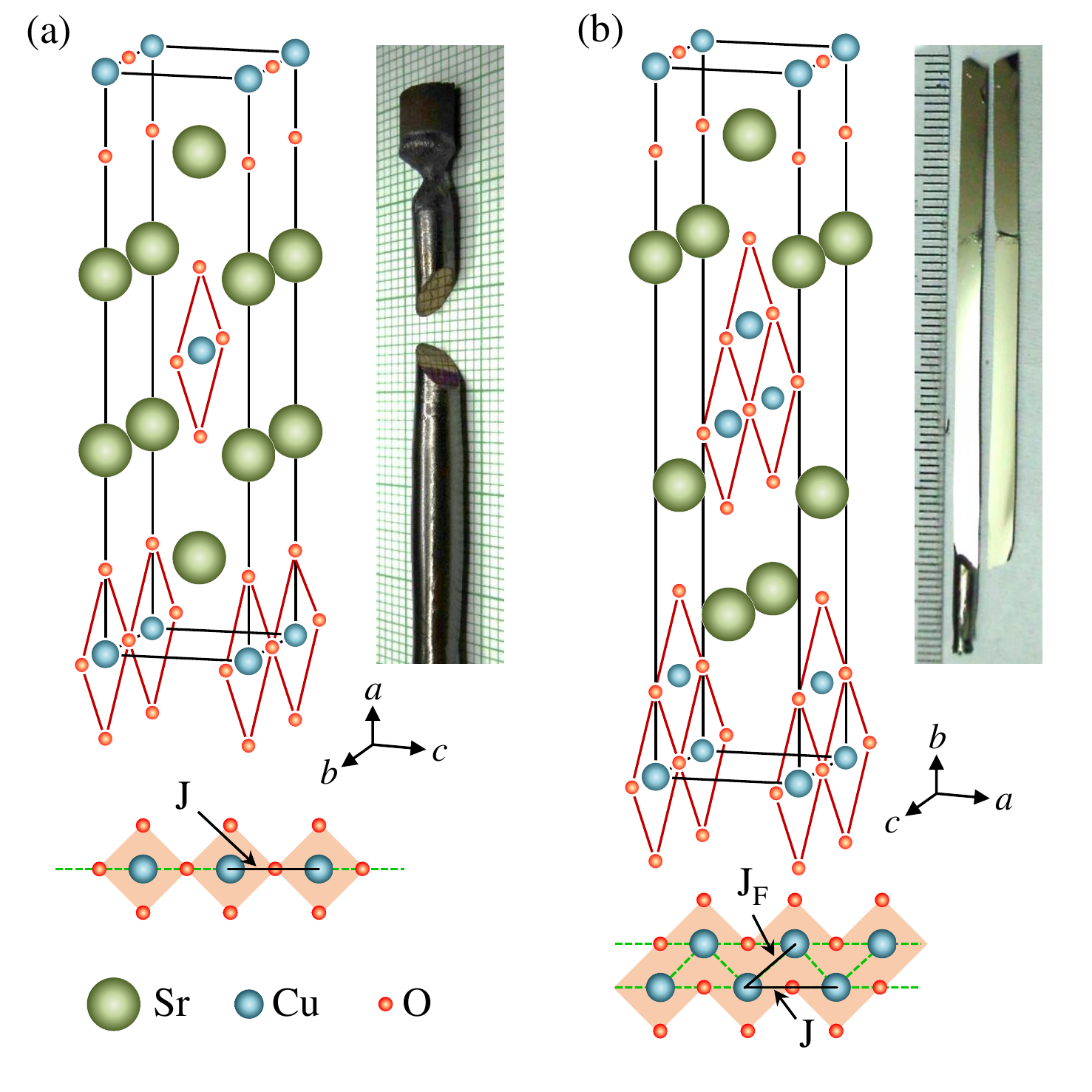}
\caption{Unit cell of (a) Sr$_{2}$CuO$_{3}$ and (b) SrCuO$_{2}$; the crystallographic directions \textit{a}, \textit{b}, and \textit{c} are shown. Representative images of the crystals are shown alongside. Lower panels show (a) the linear and (b) the zigzag chains.}

\label{Cry_str}
\end{figure}

Sr$_{2}$CuO$_{3}$ and SrCuO$_{2}$ are known to melt non-congruently. Therefore, the crystals of these compounds and their doped variants were grown using the traveling-solvent-floating-zone (TSFZ) technique associated with an image furnace. High purity precursor materials (purity $>$ 99.99\%) were used in the fabrication of these crystals, which, combined with the fact that TSFZ is a crucible free technique, resulted in high purity single crystals, which minimized the effect of extrinsic magnetic impurities. Representative images of some of the grown crystals with mirror-finished cleaved surfaces are shown in Fig. 1. This being a crucible-free technique, crystals grown using this process are generally of very high-purity, free from contamination from the flux or the crucible material. The growth details of SrCuO$_{2}$ were recently reported by us in ref. \citenum{Karmakar13}. The crystals of Sr$_{2}$CuO$_{3}$ and Sr$_{2}$Cu$_{0.99}M_{0.01}$O$_{3}$ ($M =$ Zn, Ni) were grown analogously and their details will be reported elsewhere. Here, we briefly present the growth conditions and some crucial observations pertaining to the segregation of dopants in these crystals.  

\begin{table}
\caption{Lattice parameters of the grown crystals}
\label{LP}
\begin{ruledtabular}
\begin{tabular}{l c c c c}
Sample & \textit{a} & \textit{b} & \textit{c} & Cell volume \\ 
\hline
Sr$_{2}$CuO$_{3}$ & 12.709 & 3.914 & 3.499 & 174.0\\
Sr$_{2}$Cu$_{0.99}$Zn$_{0.01}$O$_{3}$ & 12.708 & 3.913 & 3.499 & 174.0\\
Sr$_{2}$Cu$_{0.99}$Ni$_{0.01}$O$_{3}$ & 12.711 & 3.905 & 3.498 & 173.7\\
SrCuO$_{2}$ & 3.574 & 16.332 & 3.911 & 228.3 \\ 
\end{tabular}
\end{ruledtabular}
\end{table}

The crystals were grown under a flowing O$_{2}$ atmosphere at a typical growth speed of 1 mm/h. The feed and the seed rods were rotated in the opposite directions at a typical speed of 15--20 rpm to ensure temperature/chemical homogeneity of the floating zone. In the growth experiments of the doped crystals we investigated the composition of the "final" floating zone, i.e., the floating zone at the end of an experiment that had run successfully for several days yielding about 7--8 cm of grown crystal. The composition of the final floating zone was obtained as follows. The final zone was solidified by quickly decreasing the furnace power. The solidified zone was sectioned along its length (i.e., parallel to the crystal length) and polished to obtain mirror finished surfaces. These surfaces were examined using SEM and the average composition was determined using the EDX probe by performing measurements over eight distinct areas spanning the entire zone area. In the growth experiment of Sr$_{2}$Cu$_{0.99}$Zn$_{0.01}$O$_{3}$, the final zone composition contained 11.1 $\pm$ 3.5 at. \% of Zn, which is significantly larger than expected. On the other hand, no trace of Ni accumulation was detected in the corresponding Ni-doped growth experiment. The accumulation of solute in the floating zone typically arises when the segregation coefficient of solute, defined as $k = c_{c}/c_{m}$ where $c_{c/m}$ is the concentration of solute in the crystal/melt, happens to be less than 1. In  such cases, the concentration of the solute in the crystal varies along its length and can be considerably smaller than the nominal value \cite{Pfann14}. From our observation of Zn accumulation in the floating zone it is inferred that the Zn concentration in our grown crystal is smaller than the nominal 1\% in the starting material. Similar behavior has been reported in the floating-zone growth of Zn doped CuGeO$_{3}$ crystal \cite{Revcolevschi36}. The resolution limit of the EDX probe (about 1--2 at. \%) does not allow us to estimate the doping concentration in the grown crystals accurately. However, as shown in the later part of our manuscript, chain-break concentration obtained from susceptibility analysis also points at a significantly smaller Zn doping than nominal 1\% in the grown crystal.

The power x-ray diffraction shows that all the grown crystals were phase pure crystallizing with the orthorhombic space group \textit{Immm} for Sr$_{2}$CuO$_{3}$ and its doped variants, and \textit{Cmcm} for SrCuO$_{2}$. The lattice parameters of all the investigated crystals are summarized in Table \ref{LP}.

\begin{figure}
% \end{figure}
\includegraphics[scale=1]{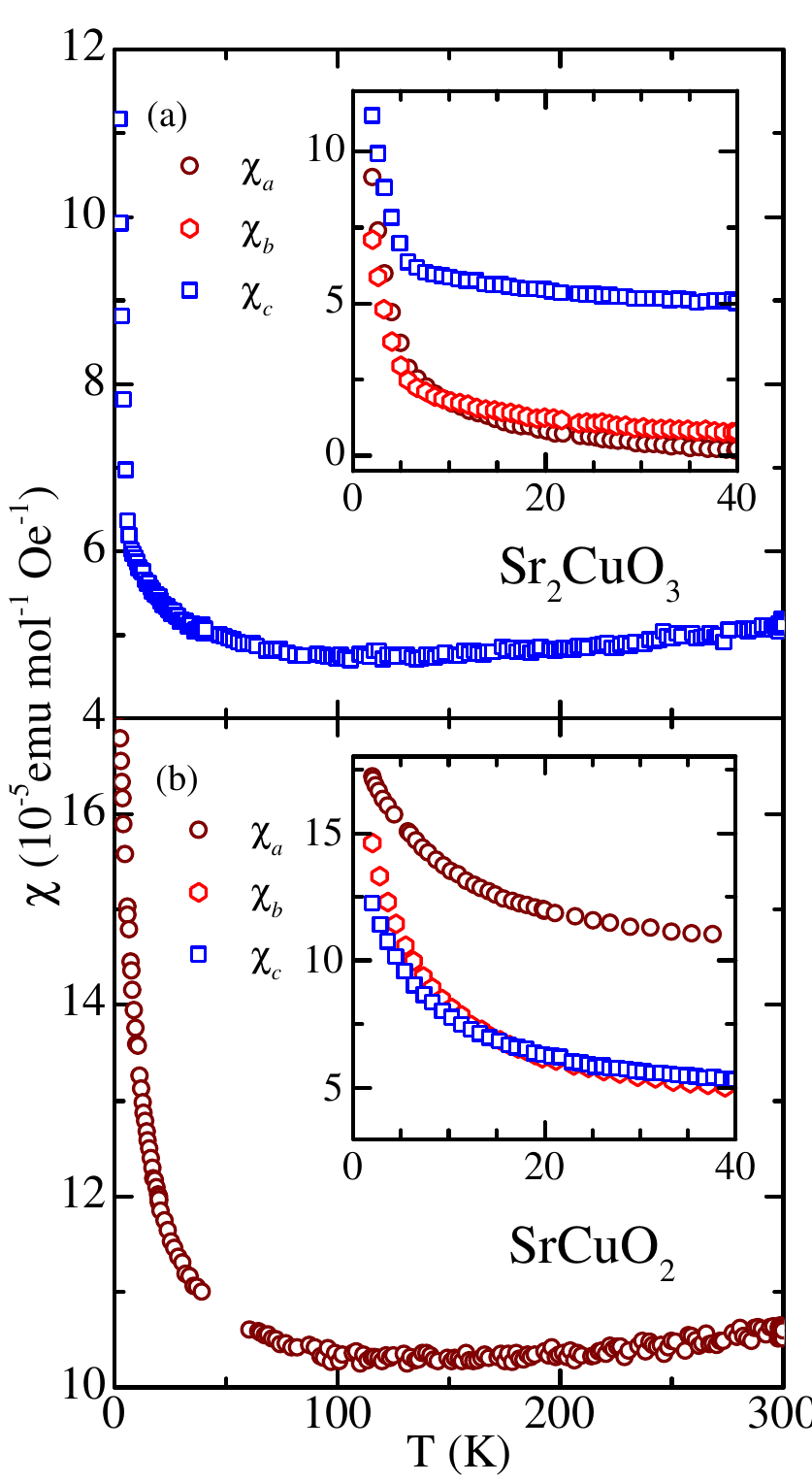}
\caption{Temperature variation of magnetic susceptibility $\chi$: (a) Sr$_{2}$CuO$_{3}$ and (b) SrCuO$_{2}$. In both cases data shown are recorded by applying field perpendicular to the CuO$_{2}$ plaquettes. Inset: low temperature susceptibilities along crystallographic \textit{a}, \textit{b}, and \textit{c} axes.}
\label{XT_Pure}
\end{figure}

The crystal structures are shown in Fig. \ref{Cry_str}. Sr$_{2}$CuO$_{3}$ consists of linear -Cu-O-Cu- chains running parallel to the crystallographic \textit{b} axis. In SrCuO$_{2}$, the chain geometry is zigzag, which can be viewed as a double-chain parallel to the \textit{c} axis that are coupled via a 90$^{\circ}$ -Cu-O-Cu- bond. The intrachain antiferromagnetic interaction (J) mediated via the linear -Cu-O-Cu- bond in both these compounds is estimated to be of the order of 2000 K. In the zigzag chains, the additional intrachain ferromagnetic J$_{F}$ via 90$^{\circ}$ -Cu-O-Cu- bond, was estimated to be about 0.1J \cite{Rice15}. While the exact role of J$_{F}$ on magnetic and thermodynamic properties is not very clear, probably due to its relatively weak strength compared to J, it seems to have no significant influence on the broad features of the spin excitation spectrum which shows the expected two-spinon gapless continuum at T = 10 K\cite{Zaliznyak16}. These observations combined with the fact that the interchain interaction (J') is rather weak in both these compounds make them excellent realizations of the 1D $S=\frac{1}{2}$ HAF model.

\section{Results and Discussions}

Figure \ref{XT_Pure} shows the low-temperature magnetic susceptibilities of the pristine Sr$_{2}$CuO$_{3}$ and SrCuO$_{2}$ crystals. The measurements were carried out under an applied magnetic field of H = 10 kOe. The data show good agreement with a previous report on well-annealed single crystals due to Motoyama \textit{et al.} \cite{Motoyama08}. The small temperature-independent anisotropy above $T \approx 20$ K is also in agreement with the previous report and is attributed to the Van Vleck paramagnetism \cite{Motoyama08}. At lower temperatures the presence of interchain interaction alters this behavior considerably. In Sr$_{2}$CuO$_{3}$, for instance, a distinct change of slope of $\chi(T)$ is observed upon cooling below $T \approx 5$ K, which coincides with the onset of long-range spin ordering in this compound \cite{Keren17}. In SrCuO$_{2}$, the absence of a similar behavior is in agreement with previous susceptibility and specific heat investigations \cite{Matsuda32}. Due to interchain interactions, the requirement of good one dimensionality breaks down at low temperature. Judging from the susceptibility anisotropy which becomes temperature dependent upon cooling below T = 20 K, we expect the interchain interactions to be important below this temperature. Since these interactions are not included in our theoretical model, therefore, in analysing the susceptibility for defect concentration analysis we considered only the data above T = 20 K. 

As seen in the main panels of Fig. \ref{XT_Pure}, upon cooling below room temperature, susceptibilities of both the compounds decrease gradually as expected since at room temperature the sample is already cooled far below the temperature $T = 0.675J/k_{B}$ where $\chi(T)$ is expected to show a broad susceptibility peak \cite{Bonner02, Griffiths04}. However, in the temperature range below about 100 K or so, the susceptibilities of both the compounds show a significant "Curie-tail" arising primarily from the chain breaks present due to slight oxygen excess in the crystals \cite{Ami07, Motoyama08}. In the presence of chain breaks one can decompose the experimentally measured susceptibility as follows: $\chi=\chi_{VV}+\chi _{core}+\chi_{Curie}+\chi _{chain}$,  where $\chi_{VV}$  is the Van-Vleck paramagnetic contribution and $\chi_{core}$  is the diamagnetic contribution of the ionic cores: both these terms are small and nearly temperature independent. Henceforth, we will use the symbol $\chi_{0}$  to represent the algebraic sum of these two terms; $\chi_{Curie}$  is the contribution arising due to the chain breaks as discussed above and $\chi_{chain}$  is the susceptibility of an uniform endless $S =\frac{1}{2}$ HAF chain in the absence of defects. Statistically, due to chain-breaks there will be as many odd-length segments as even. To determine the chain break concentration (p), one can assume that at sufficiently low temperatures, the contribution of the even-length segments dies out and that the odd-length segments, having concentration p/2, lock into a spin-doublet ground state ($S_{Z} =\pm\frac{1}{2}$) due the nearest neighbor AFM interaction \cite{Ami07}. By fitting the low temperature data using the Curie-Weiss (CW) law one can therefore get the value of p. However, this naive approach leads to a considerable underestimation of the impurity concentration \cite{Sirker09}. Moreover , in real systems, a satisfactory fit to the data can only be obtained by including the Curie temperature in the fitting expression. However, the physical significance of the Curie-temperature remains ambiguous \cite{Ami07, Kojima18}. Fujimoto and Eggert pointed out that in the presence of chain breaks, besides the 1/T contribution due to odd-length segments, one should also take into account the boundary contribution arising from the staggered moment at the free ends of a segment which also corresponds to a Curie-like behavior with logarithmic corrections \cite{Fujimoto19}. Following their work, Sirker \textit{et al.} obtained a formula for fitting the experimental susceptibility of a $S =\frac{1}{2}$ HAF chain in the presence of chain breaks including the boundary corrections \cite{Sirker12}. The equation for the average susceptibility obtained by them is as follows:

\begin{widetext}
\begin{equation}
\chi_{p}=\frac{p}{4T}\frac{(1-p)}{(2-p)}(1-(1-p)^{J/T})+(1-p)^{J/T}((1-p+\frac{pJ}{T})\chi_{chain}+p\chi_{B})
\label{XP}
\end{equation}
\end{widetext}

Here, p is the chain-break concentration and the factor $(1-p)^{J/T}$ facilitates the crossover between high- and low-temperature limits, such that, for a given value of p and J, $\chi_{p}$  tends to a value $\frac{p(1-p)}{4T(2-p)}$  for $\frac{T}{J}\ll p$  ; i.e., at very low temperatures. In this limit, $\chi_{p}$ varies as 1/T with an effective paramagnetic Curie constant $\frac{p(1-p)}{4(2-p)}\approx\frac{p}{8}$  which amounts to an effective spin $\frac{1}{2}$ paramagnetic concentration $N_{S} = \frac{p}{2}$. This is the same old result obtained naively by considering only the odd-length segments contributing to the Curie tail. At the other extreme, i.e., $\frac{T}{J}\gg p$  (which is easily reached at temperatures well below J for small p), the first term diminishes quickly to zero; however, the Curie behavior persists due to the presence of term $p\chi_{B}=p/[12Tln(2.9J/T)]$ which corresponds to the boundary susceptibility\cite{Fujimoto19}. In this case $N_{S}$ has a logarithmic temperature dependence given by $p/[3Tln(2.9J/T)]$. At any intermediate temperature the effective paramagnetic concentration can be calculated using Eq. 7.3 of Ref. \cite{Sirker12}. It should be noted that the factor $\frac{N_{A}g^{2}\mu_{B}^{2}}{k_{B}}$   (here $N_{A}$ is the Avogadro's constant, g represents the Land\'e-g factor, $\mu_{B}$ is the Bohr magneton, and $k_{B}$ is the Boltzmann constant) is not explicitly shown in the equation above for $\chi_{p}$. In previous studies, the susceptibility of these compounds is analyzed by taking a g value close to 2\cite{Ami07,Sirker09,Mahajan41}, since the spin associated with the Cu$^{2+}$ ion behaves like an isotropic Heisenberg spin, which is also reflected in the magnetic susceptibilities that show a nearly temperature-independent anisotropy. The ESR experiments \cite{Ohta42,Asakawa44} also suggest that the value of g for these compounds is not significantly different from 2. In view of this, in analyzing the susceptibility data using Eq. (\ref{XP}) we fixed the g value at 2. Slight departures from this value will not change our results significantly.

$\chi_{chain}$ in Eq. (\ref{XP}) is replaced by the fit function obtained by Johnston \textit{et al.} [Ref. \cite{Johnston21}, fit 1, Eq. (50)]
\begin{equation}
\chi_{chain} =\frac{1}{4T}\Bigg[ \frac{1+\sum\nolimits_{n=1}^q \frac{N_{n}}{(T/J)^{n}}}{1+\sum\nolimits_{n=1}^r \frac{D_{n}}{(T/J)^{n}}}\Bigg]
\end{equation}
The values of N$^{n}$’s and D$_{n}$'s are taken from ref. \cite{Johnston21}. The expression yields highly accurate values of $\chi_{chain}$ in the temperature range 0.01J $<$ T $<$ 5J). Since, J for the compounds under consideration is of the order of 2000 K, this expression is quite appropriate in the temperature range over which the data has been analyzed in the present work. It is worth mentioning here that the numerical result for $\chi_{chain}$ due to BF \cite{Bonner02} which is frequently used in the literature to fit the experimental data of spin chains is accurate only at high temperatures $T \geq 0.4 J/k_{B}$ which makes its applicability inappropriate in the present case. This non-applicability of the BF model in fitting the susceptibilities of these compounds has been demonstrated by Motoyama \textit{et al.} \cite{Motoyama08}.  

\textbf{\textit{Fitting procedure}:} In Eq. (\ref{XP}), there are three fitting parameters: (1) the defect concentration p, (2) the intrachain exchange J and (3) the temperature independent contribution to the susceptibility, $\chi_{0}$ ($\chi_{p} = \chi - \chi_{0}$). Fitting the experimental data to Eq. (\ref{XP}) can give rise to arbitrary results if the parameters are not initialized judiciously. This problem is further compounded by the fact that in the temperature range of our measurements the susceptibility values are rather small ($\approx 5 \times 10^{-5}$ emu/mol.Oe). In order to avoid these difficulties, we determined these parameters in steps by selecting suitable temperature ranges where Eq. (\ref{XP}) can be further simplified and the number of parameters reduced. 

In the first step, we considered a temperature range over which the measured susceptibility is largely 'p' independent which allowed for the estimation of $\chi_{0}$ and J. For example, in the high temperature range $250 K \leq T \leq 300 K$ of our measurements the contribution of the first term ($\approx p / 8T$, where $p \leq 0.01$ ) is negligible. This leaves us with only the $\chi_{chain}$  and the $\chi_{B}$  terms to deal with in Eq. \ref{XP}. We further note that the relative magnitude of $p\chi_{B}$   is very small compared to $(1-p+pJ/T)\chi _{chain}$. For instance, if we assume p to have a value of 0.005 and take J to be around 2000 K then near $T = 300 K$ the value of the term $p\chi _{B}\approx7\times10^{-7}$emu/mol.Oe  and that of  $(1-p+pJ/T)\chi _{chain}\approx9\times10^{-5}$emu/mol.Oe, where the value of $\chi_{chain}$ itself is $\approx8.7\times10^{-5}$ emu/mol.Oe which is almost 97 \% of $(1-p+pJ/T)\chi _{chain}$. Therefore, by taking the initial values of J and $\chi_{0}$  from ref. \cite{Motoyama08}, we fitted the measured susceptibility in the above temperature range simply by: $\chi=\chi_{0}+\chi_{chain}$. This gave us a fairly good estimate on the values of $\chi_{0}$ and J. In the next step, we used these values to fit the data from 20 to 300 K using Eq. (\ref{XP}) by allowing p to vary. In the final step, all three parameters were varied around their values obtained in the previous steps to estimate the size of the error bars.

\begin{figure}
\includegraphics[scale=1]{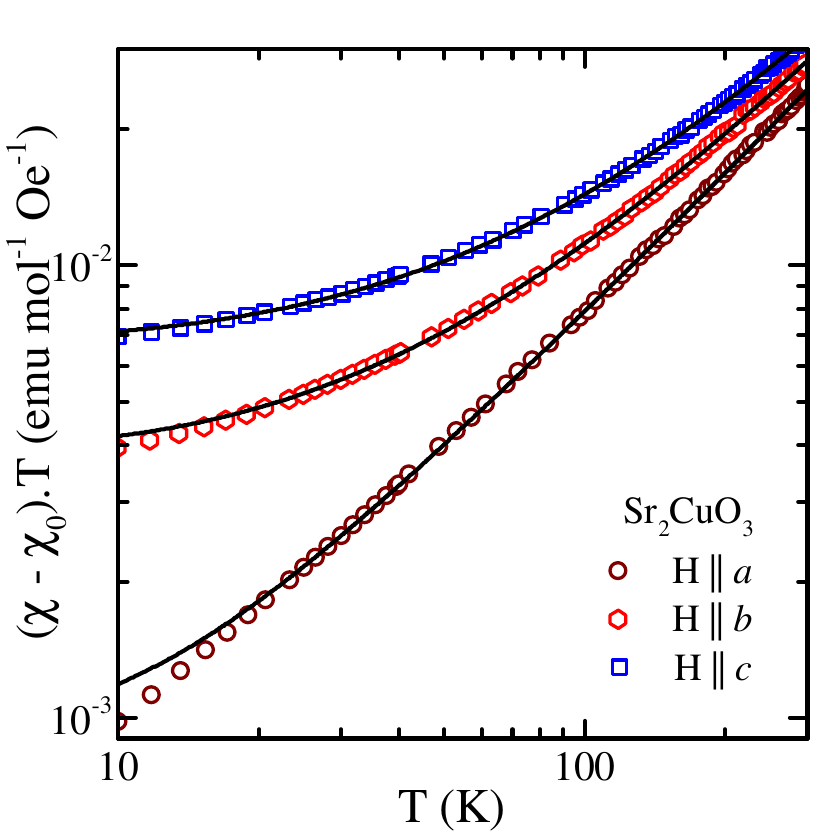}
\caption{Magnetic susceptibility $(\chi-\chi_{0})T$ of Sr$_{2}$CuO$_{3}$. Solid lines represent the theoretical fit using Eq. (\ref{XP}) (see text for details). Data along the \textit{b} and the \textit{c} axes are shifted by multiples of 3 x 10$^{-3}$.}
\label{XP_pure}
\end{figure}

\begin{table*}
\caption{Values of intra-chain exchange J, the temperature independent contribution $\chi_{0}$ and defect concentration p obtained by fitting the experimental susceptibility using Eq. (\ref{XP}). For comparison, the low-temperature data were also fitted to the Curie-Weiss equation. $p\_{CW}$ and $\theta_{P}$ are the fitting parameters of this fit (see text for details).}
\begin{ruledtabular}
\begin{tabular}{l c c c c c c}
\multicolumn{2}{p{4cm}}{\multirow{2}{*}{Parameters}} & J & $\chi_{0}$ & p & p$_{CW}$ & $\theta_{p}$\\
 &  & ($k_{B}$ K) & (10$^{-5}$emu/mol.Oe) & (per Cu site) & (per Cu site) & (K)\\ 
\hline
\multirow{3}{*}{Sr$_{2}$CuO$_{3}$} & (H $\parallel$ a) & \multirow{3}{*}{2100$\pm$200} & -8.0$\pm$0.2 & 0.005 & 0.001 & 2.9 \\
 & (H $\parallel$ b) & & -7.7$\pm$0.2 & 0.005 & 0.001 & -1.5 \\
 & (H $\parallel$ c) & & -3.6$\pm$0.2 & 0.004 & 0.001 & -4.8 \\
\hline
\multirow{3}{*}{SrCuO$_{2}$} & (H $\parallel$ a) & \multirow{3}{*}{1930$\pm$200} & 1.3$\pm$0.2 & 0.007 & 0.003 & -2.7 \\
 & (H $\parallel$ b) & & -4.3$\pm$0.2 & 0.008 & 0.003 & -0.7 \\
 & (H $\parallel$ c) & & -4.3$\pm$0.2 & 0.008 & 0.003 & -3.5 \\
 \hline
Sr$_{2}$Cu$_{0.99}$Zn$_{0.01}$O$_{3}$ & (H $\parallel$ c) & \multirow{2}{*}{2100$\pm$200} & -3.1$\pm$0.2 & 0.011 & 0.004 & -1.5\\
Sr$_{2}$Cu$_{0.99}$Ni$_{0.01}$O$_{3}$ & (H $\parallel$ c) & & -3.5$\pm$0.2 & 0.018 & 0.006 & 7.1\\ 
\end{tabular}
\end{ruledtabular}
\label{FP}
\end{table*}

\textbf{\textit{Susceptibility of Sr$_{2}$CuO$_{3}$ crystal}:} In Fig. \ref{XP_pure} we show the temperature variation of quantity $(\chi-\chi_{0})T$ along with the fitted curves. The data and the fitted curves are shown on a log-log plot to emphasize the low-temperature behavior. A satisfactory agreement of the experimental data with the fitted curve is observed over the entire fitting range. However, when extrapolated to lower temperatures the fitted curve tends to saturate to the value $ \frac{p (1-p)}{4 (2-p)} \approx \frac{p}{8}$, whereas the experimental data continue to decrease due to the intrachain exchange not included in our fitting equation. The p values along the three crystallographic axes are listed in table \ref{FP}. The average p value came out to be to be 0.005 per Cu site. This value is consistent with a value of 0.006 per Cu site reported by Sirker \textit{et al.} for their unannealed crystal. An average J value of $2100_{-200}^{+200}$ K and $\chi_{0}$ values (in the units of $10^{-5}$ emu/mol.Oe) of $-8.0 \pm 0.2 (\chi _{0}^{a}), -7.7 \pm 0.2 (\chi _{0}^{b})$ and $-3.6 \pm 0.2 (\chi _{0}^{c})$, where the superscript on $\chi _{0}$  denotes the crystal axis along which the external magnetic field is applied are well within the expected range based on the previous studies. For example, Motoyama et al. obtained a value of $J =2200_{-200}^{+200}$  by analyzing the susceptibility data of annealed crystals up to temperatures as high as $T = 800 K$ \cite{Motoyama08}. Here, it should be mentioned that somewhat higher values of J were reported from the optical absorption (J = 2850 K)\cite{Suzuura20} and the low-temperature specific heat data (J = 2500 K) \cite{Sologubenko37}, respectively. However, based on a Quantum Monte Carlo study it has been suggested that the value of J estimated from $\chi(T)$ data may be reduced due to phonon induced fluctuations of J not included in fitting the $\chi(T)$ data \cite{Sandvik23}.

In Table \ref{FP} we also list the chain-break concentration obtained by fitting the susceptibility data between $T = 20 K$ and $40 K$ using the CW law. To fit the CW equation, we subtracted the $\chi _{0}$ and $\chi _{chain}$ contributions from the raw data. $\chi _{chain}$ is calculated using the J value listed in Table \ref{FP}. An average chain-break concentration of 0.0012 per Cu site obtained from the CW fit agrees well with that reported previously \cite{Motoyama08,Kojima18}, where a similar method has been used. However, as expected, this value is considerably smaller than that obtained using a more rigorous fitting procedure where the boundary contributions are also taken into account.   

\begin{figure*}
\includegraphics[scale=1]{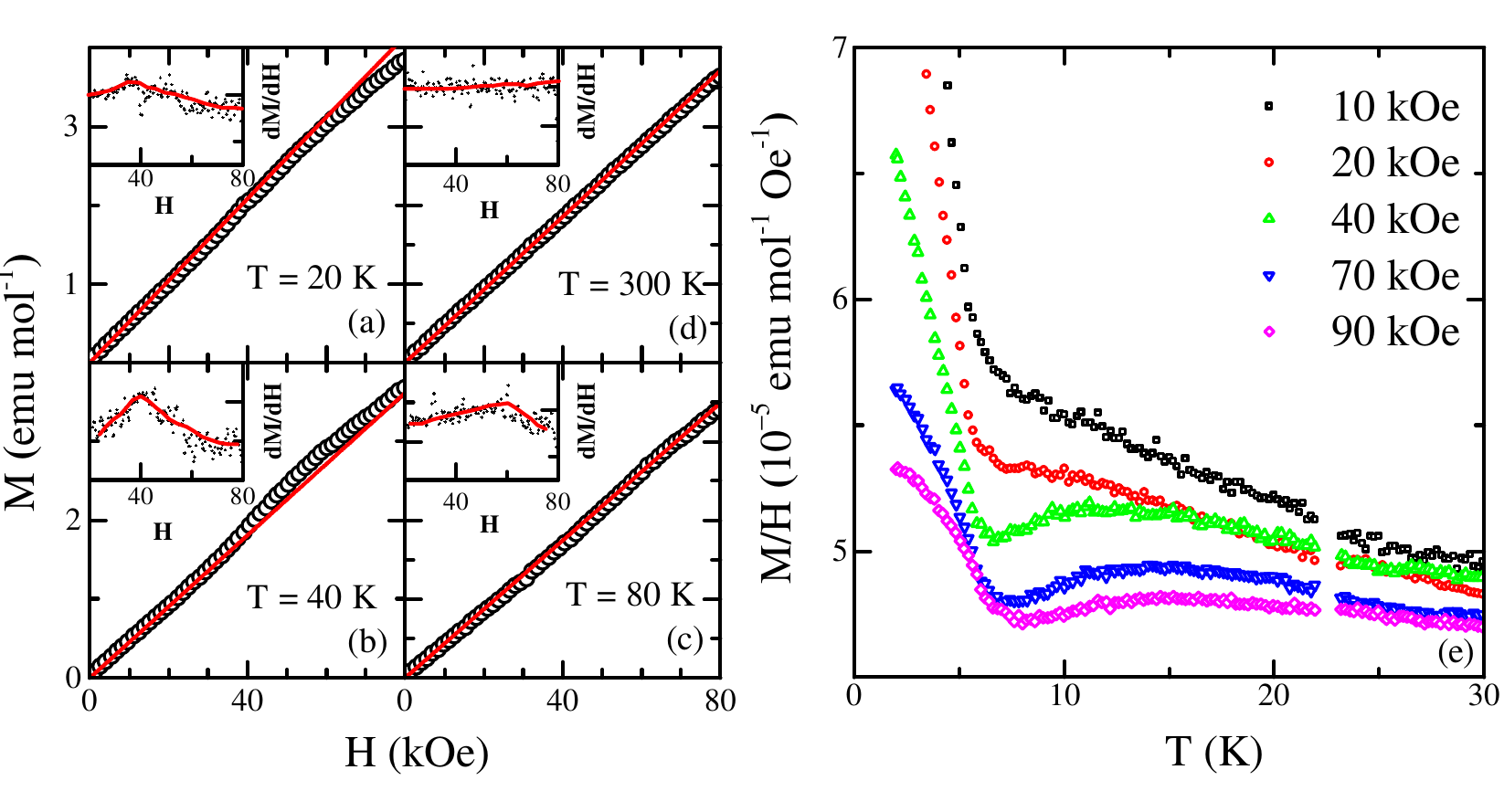}
\caption{(a)--(d) Isothermal magnetization (M) of Sr$_{2}$CuO$_{3}$ shown as function of applied magnetic field (H). Solid curves are calculated M(H) (see text for details). Inset: derivative (dM/dH) as a function of H. Lines are guide to the eye; (e) temperature dependence of M/H under various applied fields. In each panel data are shown for field applied perpendicular to the CuO$_{2}$ plaquettes}
\label{MH_pure}
\end{figure*}

As noted earlier, the effective paramagnetic concentration ($N_{S}$) in our model is a temperature dependent quantity that shows a non-monotonic behavior, decreasing rapidly upon increasing the temperature (from a value of $N_{S}= p / 2$ at $T = 0$) before increasing again at high temperatures where it goes as $p ⁄ [3ln(2.9J/T)$ ] . At any intermediate temperature T, $N_{S}$ can be estimated from $\chi_{p}(T)$ by subtracting from it the chain susceptibility $\chi_{chain}(T)$ and then by equating the temperature times this difference to an effective temperature dependent Curie constant $C(T)$ (see Eq. 7.3 of Ref. \cite{Sirker12}). Using C(T), the value of $N_{S}$ can be readily calculated (see, for example, Ref. \citenum{Ashcroft31}), using C = (N$_{s}$N$_{A}$g$^{2}\mu_{B}^{2}$S(S+1))/3k$_{B}$. In table \ref{NS}, we list the values of $N_{S}$ for our Sr$_{2}$CuO$_{3}$ crystals at temperatures $T = 20, 40, 80, 200$ and $300 K$.  In Fig. \ref{MH_pure}, we compare the calculated M(H) consisting of a Brillouin term accounting for the $S = \frac{1}{2}$ paramagnetic impurities (numbering $N_{S}$ per Cu mol) and the terms linear in H consisting of the contribution due to $\chi _{0}$ and $\chi _{chain}$. The calculated curves show a good overall agreement with the experimental data at all temperatures and upto the highest applied field. The experimental data, however, show a weak metamagneticlike transition which is evident from a peak in the $dM/dH$ plots shown in the insets. The peak position shifts to higher field values with increasing temperature, falling beyond the measurement range of our experimental setup at T = 300 K, where M(H) shows a linear behavior upto H = 80 kOe, in good agreement with the calculated curve. We shall continue our discussion of this field dependence below. Here, to conclude this section, we note that the values of the fitting parameters obtained from our $\chi(T)$ analysis provide a good overall description of the isothermal magnetization at different temperatures that span a fairly wide range thereby reinforcing confidence in these values. In the future, it will be interesting to investigate how this weak metamagneticlike feature seen in M(H) affects the quality of $\chi$(T) fit at low-temperatures.

\begin{table}
\centering
\caption{Effective paramagnetic impurities in the pristine Sr$_{2}$CuO$_{3}$.}
\begin{ruledtabular}
\begin{tabular}{c c}
T (K) & $N_{S}$ ($10^{-4}$per Cu site)\\
\hline
20 & 6.7\\
40	 & 5.1\\
80 & 4.5\\
300 & 5.3\\
\end{tabular}
\end{ruledtabular}
\label{NS}
\end{table}

\textbf{\textit{Field dependence of magnetization in Sr$_{2}$CuO$_{3}$}:} To probe the observed metamagneticlike behavior (Fig. \ref{MH_pure}), we measured the temperature variation of magnetization under constant applied magnetic fields of increasing strength. The resulting data, plotted as M/H against T, is shown in the panel (e) of Fig. \ref{MH_pure}. Under low fields (H = 10 kOe), the data show a steep increase below $T = 5 K$ which coincides with the magnetic ordering temperature. At higher fields ($H \approx 20 kOe$ and higher), a broad maximum above the ordering temperatures is observed, whose position shifts to higher temperatures with increasing field strength. Given that the intrachain coupling is around 2000 K, the observed field dependence of the chain magnetization under relatively small applied magnetic fields appears surprising. This behavior is confirmed by performing similar measurements along the other two crystallographic directions (not shown here). While it is difficult to pin down the origin of this behavior, it is interesting to note that $^{63}$Cu NMR spectra recorded on a high-quality crystal of Sr$_{2}$CuO$_{3}$ also exhibit anomalous features in the same field-temperature range \cite{Takigawa25}. The field sweep NMR spectra develop a broad background with sharp edges at low temperatures, which was theoretically predicted as arising due to external magnetic field induced local staggered magnetization near the chain breaks \cite{Eggert11}. However, two additional features in the NMR spectra, not expected from the theory of Ref. \cite{Eggert11}, were also observed, viz., splitting of the main NMR peak and appearance of shoulders on both sides of the main peak. In one of the theoretical studies, these features were attributed to "mobile bond defects," which can arise due to spin-lattice coupling and result in a "local alternating magnetization"  \cite{Boucher26}. In Ref. \citenum{Sirker27}, however, the role of interchain exchange  in giving rise to these additional NMR features has been scrutinized. In the future, it will be interesting to investigate if there is any correspondence  between the anomalous NMR line-shape and magnetization behavior observed in our experiments. 

\begin{figure}
\includegraphics[scale=1]{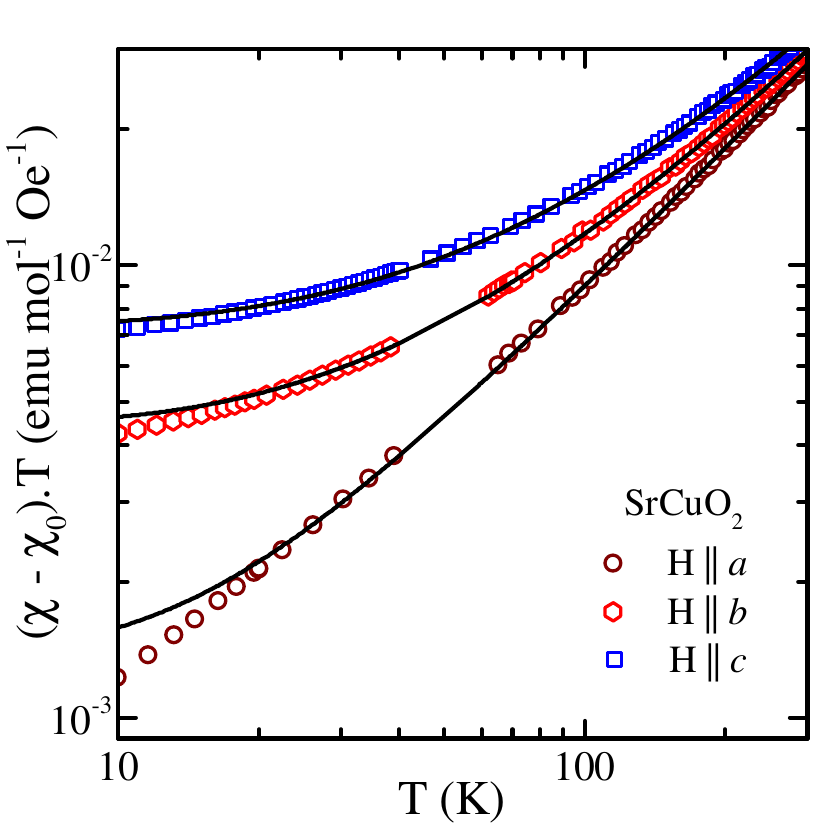}
\caption{Magnetic susceptibility $(\chi-\chi_{0})T$ of SrCuO$_{2}$. Solid lines represent the theoretical fit using Eq. \ref{XP} (see text for details). Data along \textit{b} and \textit{c} axes are shifted by multiples of 3 x 10$^{-3}$}
\label{XP_112}
\end{figure}

\textbf{\textit{Susceptibility of SrCuO$_{2}$ crystals}:} Next, we fitted the susceptibility of SrCuO$_{2}$.  The fitted curve is shown in Fig. \ref{XP_112}. A satisfactory fit to the data is obtained down to nearly T = 30 K. An average J value of $1930_{-200}^{+200} K$ and $\chi_{0}=1.3 \pm 0.2 (\chi_{0}^{a}), -4.3 \pm 0.2 (\chi_{0}^{b}), -4.3 \pm 0.2 (\chi_{0}^{c})$ (in the units of 10$^{-5}$ emu/mol.Oe) obtained from the fitting analysis are in good agreement with Motoyama et al.\cite{Motoyama08}. The average defect concentration turned out to be 0.008 per Cu site which is higher than that in Sr$_{2}$CuO$_{3}$ which was grown and annealed under similar conditions but prepared using starting precursors with slightly higher purity. 

\begin{figure}
\includegraphics[scale=1]{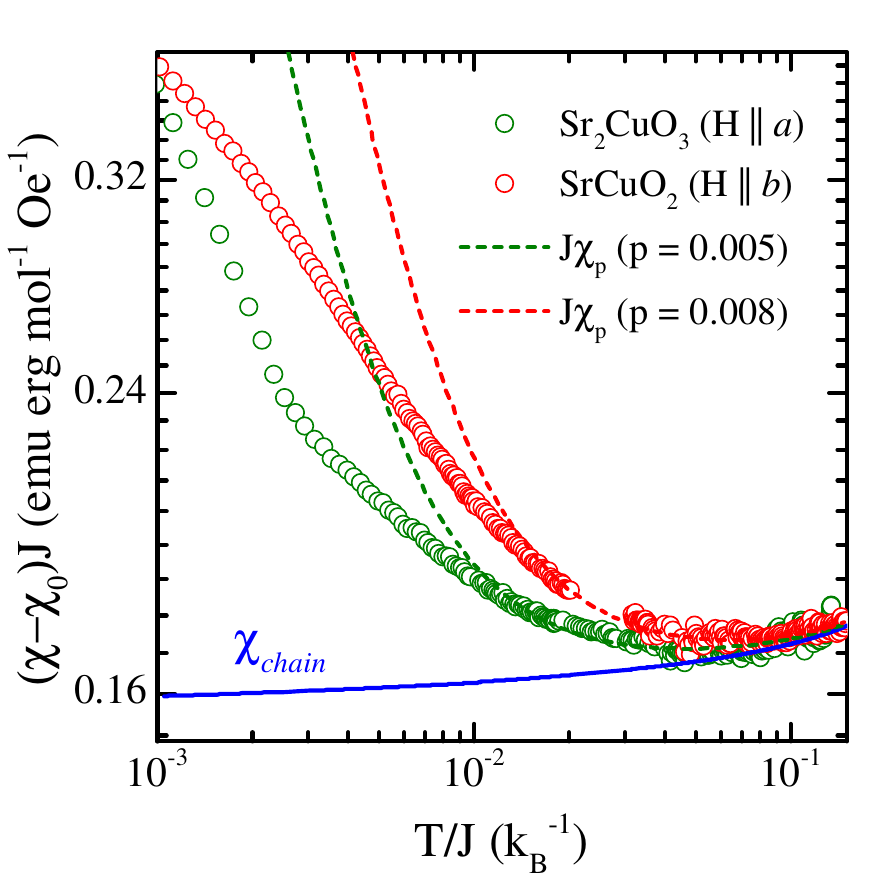}
\caption{Normalized susceptibilities of Sr$_{2}$CuO$_{3}$ and SrCuO$_{2}$. In both cases data shown are for field applied along the longest crystallographic axis. The calculated chain susceptibility $\chi_{chain}J$ is also shown (solid line). The dashed lines represent calculated $\chi_{p}J$ for p = 0.005 and 0.008 per Cu site (see text for details).}
\label{SCO_LCC}
\end{figure}

In Fig. \ref{SCO_LCC} we compare the susceptibilities of the two compounds by plotting $(\chi - \chi_{0})J$ against T/J where the values of $\chi_{0}$ and J for the two compounds are used from table \ref{FP}. In this representation, the susceptibility is J-independent. On the same plot we also show the variation of $\chi_{chain}J$ \cite{Lukyanov06}. Near room temperature the two data sets collapse on the calculated $\chi_{chain}$ as one would have expected. However, upon lowering the temperature below $T \approx 0.1 J$ the measured susceptibilities start to deviate from the ideal chain behavior: increasing with decreasing temperature due to chain breaks. This increase is more pronounced in SrCuO$_{2}$ than in Sr$_{2}$CuO$_{3}$, in agreement with the higher p value for the former. The double chains of SrCuO$_{2}$ are characterized by the presence of an additional ferromagnetic frustrated exchange $J_{F}$ which is estimated to about $0.1 J$ \cite{Rice15}. While the dominant antiferromagnetic exchange (J) tends to lower the chain susceptibility upon lowering the temperature, $J_{F}$ has just the opposite effect as it tends to disrupt any tendency towards antiferromagnetic ordering. Therefore, the chain susceptibility $\chi_{chain}$ in SrCuO$_{2}$ is expected to show a slower rate of decrease upon cooling than for a linear chain as was used in Eq. (\ref{XP}) for the estimation of p. Also,  the presence of coupling J$_{F}$ in the double chains itself implies that a defect cannot be considered as a chain break at low temperatures. Therefore, at temperatures below the rung coupling J$_{F}$, Eq. (\ref{XP}) has only a limited validity. The p value obtained here for undoped SrCuO$_{2}$ is, therefore, possibly slightly overestimated. A more direct manifestation of J$_{F}$ evidently is to suppress the long-range spin ordering which has been studied previously using $\mu$SR and neutron studies \cite{Matsuda32}. It is worth mentioning here that though the next-nearest neighbor interaction $J_{nnn}$ along the chain length is estimated to be around 140 K\cite{Rosner28}, Eq. (\ref{XP}) is shown to be applicable as long as $T \gg p^{2}J_{nnn}$ \cite{Sirker12}. Thus J$_{nnn}$ is not expected to have any significant influence either on the susceptibility or on the determination of p.

\begin{figure}
\includegraphics[scale=1]{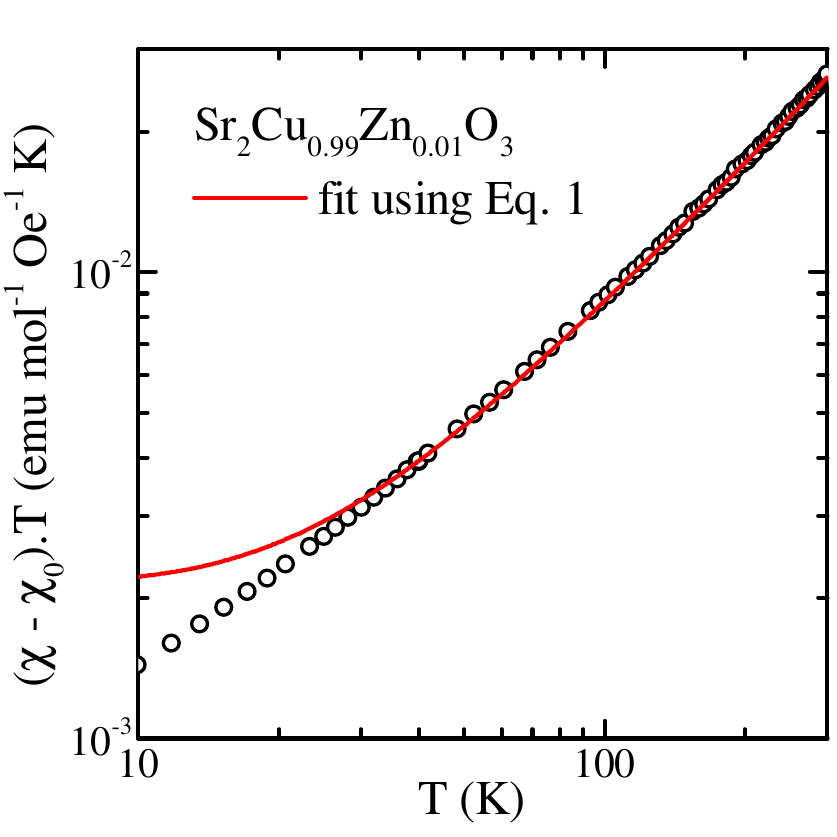}
\caption{Magnetic susceptibility $(\chi-\chi_{0})T$ of Sr$_{2}$Cu$_{0.99}$Zn$_{0.01}$O$_{3}$. Solid curve in the main panel represents the best theoretical fit using Eq. (\ref{XP})}
\label{XP_Zn1}
\end{figure}

\textbf{\textit{Susceptibility of Sr$_{2}$Cu$_{0.99}$Zn$_{0.01}$O$_{3}$}:} We investigated the behavior of Sr$_{2}$CuO$_{3}$ crystal doped with non-magnetic Zn (nominal doping level $\sim 1\%$). As mentioned in the crystal growth section, by analyzing the composition of the frozen-in floating zone we had inferred that the Zn concentration in the grown crystal should be significantly smaller than the nominal value. To test this assertion, we analyzed the susceptibility data of the Zn-doped crystals using the same procedure as used for the pristine crystals. The results are shown in Fig. \ref{XP_Zn1}. A reasonably good fit to the experimental data is observed down to nearly T = 40 K. The value of the exchange integral did not show any appreciable decrease upon doping, which is not surprising given the dilute levels of doping. Similarly, $\chi_{0}$ also did not show any significant variation with respect to the pristine compound (Table \ref{FP}). The chain-break concentration was found to be around 0.011(2) per Cu site. Taking a cue from the work of Sirker \textit{et al.}, where the chain-break concentration in the doped crystals exceeded the nominal Pd concentration by an amount roughly equal to the concentration of intrinsic chain breaks in their pristine compound, we conclude that the concentration of chain breaks induced by Zn doping in our crystal is around $\sim$ 0.006, the remaining $\sim$ 0.005 being due to the intrinsic defects as inferred from the analysis of the pristine compound. This smaller Zn concentration of 0.6\% obtained from the susceptibility analysis is in agreement with the crystal growth results.
 
\textbf{\textit{Susceptibility of Sr$_{2}$Cu$_{0.99}$Ni$_{0.01}$O$_{3}$}:} Recently, it was reported that upon 1\% Ni doping in SrCuO$_{2}$ a dramatic change of the low-energy spin excitations takes place. Briefly, what had been a gapless two-spinon excitation spectrum for the pristine compound became gapped upon Ni doping \cite{Simutis30}. The value of the spin pseudogap was found to be around $\sim 8$ meV. In our own preliminary experiments using inelastic neutron scattering we found a similar gap in Ni-doped Sr$_{2}$CuO$_{3}$. However, no such gap was found in the Zn$^{2+}$ doped crystals \cite{Karmakar38}. Eggert and Affleck \cite{Eggert10} investigated the behavior of an isolated magnetic impurity substituted in a half-odd-integer-spin HAF chain theoretically. In particular, they considered an isolated $S = 1$ impurity (e.g., Ni$^{2+}$) in the chain and showed that in the limit of infinite chain length Ni$^{2+}$ moment either remains decoupled from the chain or is completely screened by the antiferromagnetically coupled neighboring spins. In the latter case, the screened Ni$^{2+}$ moment essentially breaks the chain at the impurity site.

\begin{figure}
\includegraphics[width=8.5cm]{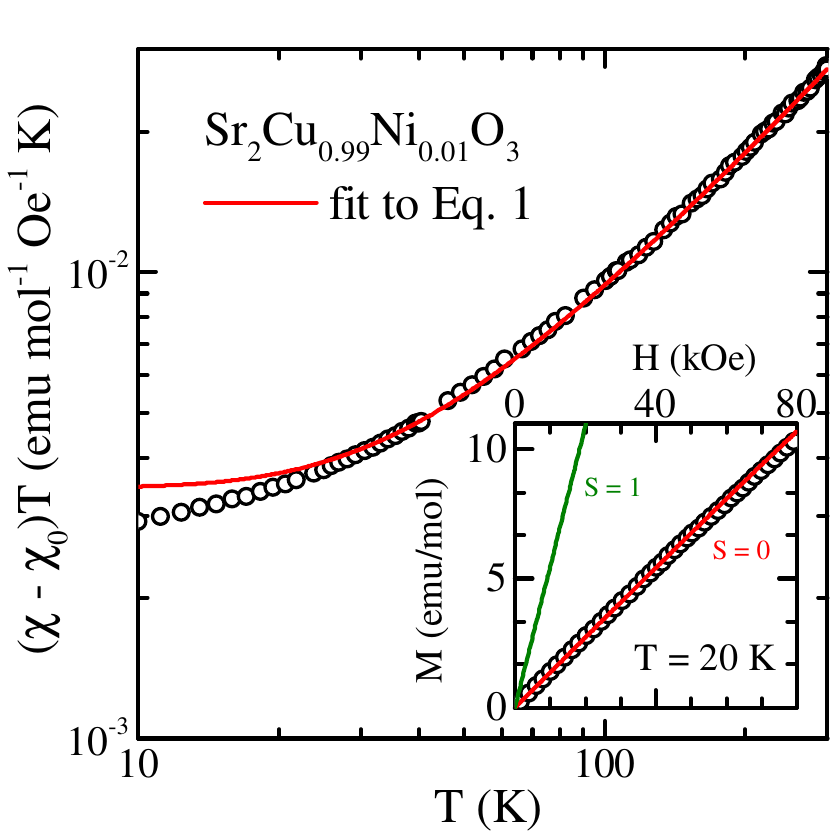}
\caption{Magnetic susceptibility $(\chi-\chi_{0})T$ of Sr$_{2}$Cu$_{0.99}$Ni$_{0.01}$O$_{3}$. Solid curve in the main panel represents the best theoretical fit using Eq. (\ref{XP}). Inset: isothermal magnetization at $T = 20$ K. Solid curves are calculated M(H) for S = 0 and S = 1 cases (see text for details).}
\label{XP_doped}
\end{figure}

Therefore, if the spin of Ni$^{2+}$ ion is indeed screened we should be able to get the Ni concentration in our crystal using Eq. (\ref{XP}). The measured susceptibility of the Ni-doped crystal was analyzed using Eq. (\ref{XP}). In Fig. \ref{XP_doped} we show the calculated susceptibility for p = 0.018(1) obtained by fitting the experimental data using Eq. (\ref{XP}) as done for the pristine crystal. Within the accuracy of our method, the fitted values of J and $\chi_{0}$ are the same as for the pristine crystal (see,  Table \ref{FP}). The defect concentration estimated here is higher than the nominal value of 0.01. However, if one takes into account the intrinsic defect concentration due to oxygen off-stoichiometry, deduced from the analysis of the pristine compound prepared and annealed under the identical conditions, we get a value of around 0.013(1), which is indeed close to the nominal concentration. We also compared the measured M(H) at $T = 20$ K with that calculated for two different spin states: S = 1 and  S = 0 for Ni concentration of 1 \%. The comparison is shown in the inset of Fig. \ref{XP_doped}. The calculated magnetization in the S = 1 case is clearly very large and deviates considerably from the experimental data. On the other hand, in the S = 0 case a close match with the experimental data is observed. There is, therefore, an overall agreement between the experimental data and the calculated curve that proves the point that Ni$^{2+}$ spins are indeed screened. This is also in agreement with the analysis reported in ref. \citenum{Mahajan41} on the powder samples of Ni doped Sr$_{2}$CuO$_{3}$, where it was shown that the Curie tail at low temperatures is considerably smaller than expected from unscreened S = 1 Ni moments.

It should be pointed out that the fit-quality for the Ni-doped case using Eq. \ref{XP} is not as good as for the pristine crystal, particularly below T = 100 K. The reason why the fitting Eq. \ref{XP} did not work so well is probably related to the fact that the ground state of Ni-doped crystal is gapped which makes the $\chi_{chain}$ used in Eq.(\ref{XP}) not suitable at temperatures below which the gap opens. In analogy with SrCuO$_{2}$, the gap value is  expected to be around 8 meV ($\approx$ 80 K) for a similar Ni concentration, which roughly coincides with the temperature below which the fit quality deteriorates. To conclude, our analysis of the magnetization data supports the screening of the Ni$^{2+}$ moments. 
         
\begin{figure}
% \end{figure}
\includegraphics[scale=1]{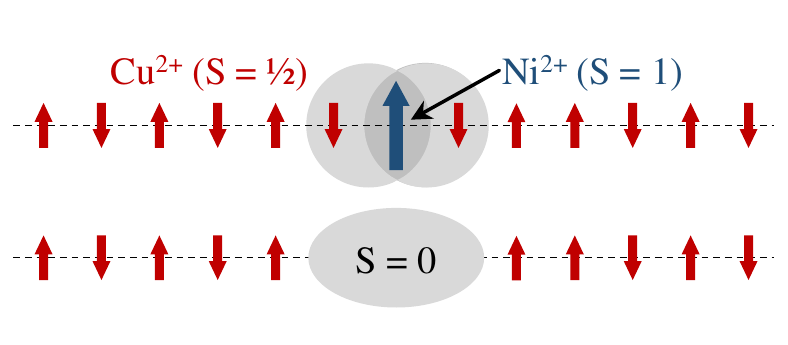}
\caption{Cartoon showing screening of the Ni spin (S = 1) coupled antiferromagnetically to it neighbors in a spin-$\frac{1}{2}$ antiferromagnetic chain.}
\label{screening}
\end{figure}

It is intriguing as to why Ni doping opens up a gap while the Zn doping apparently does not. An important difference between these two impurity types is that unlike the chain breaks due to Zn$^{2+}$ impurity, where the spins on the either side of the impurity may still couple weakly through the next-nearest-neighbor interaction, the Ni$^{2+}$ impurity results in a spin compensated cloud around the impurity site, which isolates the chain segments (Fig. \ref{screening}). In such a scenario, whose validity needs to be established through further work, the spinons will be confined over finite-length sections of the chain which should open up a gap in the low-energy spectrum with the gap value scaling inversely with the average length of the segments ($\overline{L}\sim \frac{1}{p}$). Another possible reason for not having found a gap in the Zn$^{2+}$ crystals might be related to the fact that Zn$^{2+}$ does not readily replace Cu$^{2+}$ in a crystal grown from the melt; hence the quantity of Zn doped in the chains may not be enough to open up an observable gap. Recently, spin gap has also been reported in the crystals of  Sr$_{1.9}$Ca$_{0.1}$CuO$_{3}$ \cite{Hammerath39} and Sr$_{0.9}$Ca$_{0.1}$CuO$_{2}$ \cite{Hammerath40}. Since Ca$^{2+}$ replaces Sr$^{2+}$, the -Cu-O-Cu- chains in these crystals should remain unsegmented. It has been argued that the ionic-size mismatch of Ca$^{2+}$ and Sr$^{2+}$ gives rise to a local bond disorder of the intrachain exchange coupling J which is probably responsible for the observed spin gap. The appearance of spingap is, therefore, not always associated with the segmentation of the chains and substantial bond disorder can also cause the ground state to be gapped. In the light of these recent results, it is fair to admit that the chain break picture is probably too naive. Even the notion of screening, strictly speaking, is valid at T = 0. At non-zero temperatures, the susceptibility will be affected by the screening cloud around the impurity spin, which extends beyond nearest neighbors making the applicability of Eq. (\ref{XP}) in such scenarios rather restrictive.  

\begin{figure}[h]
% \end{figure}
\includegraphics[scale=0.95]{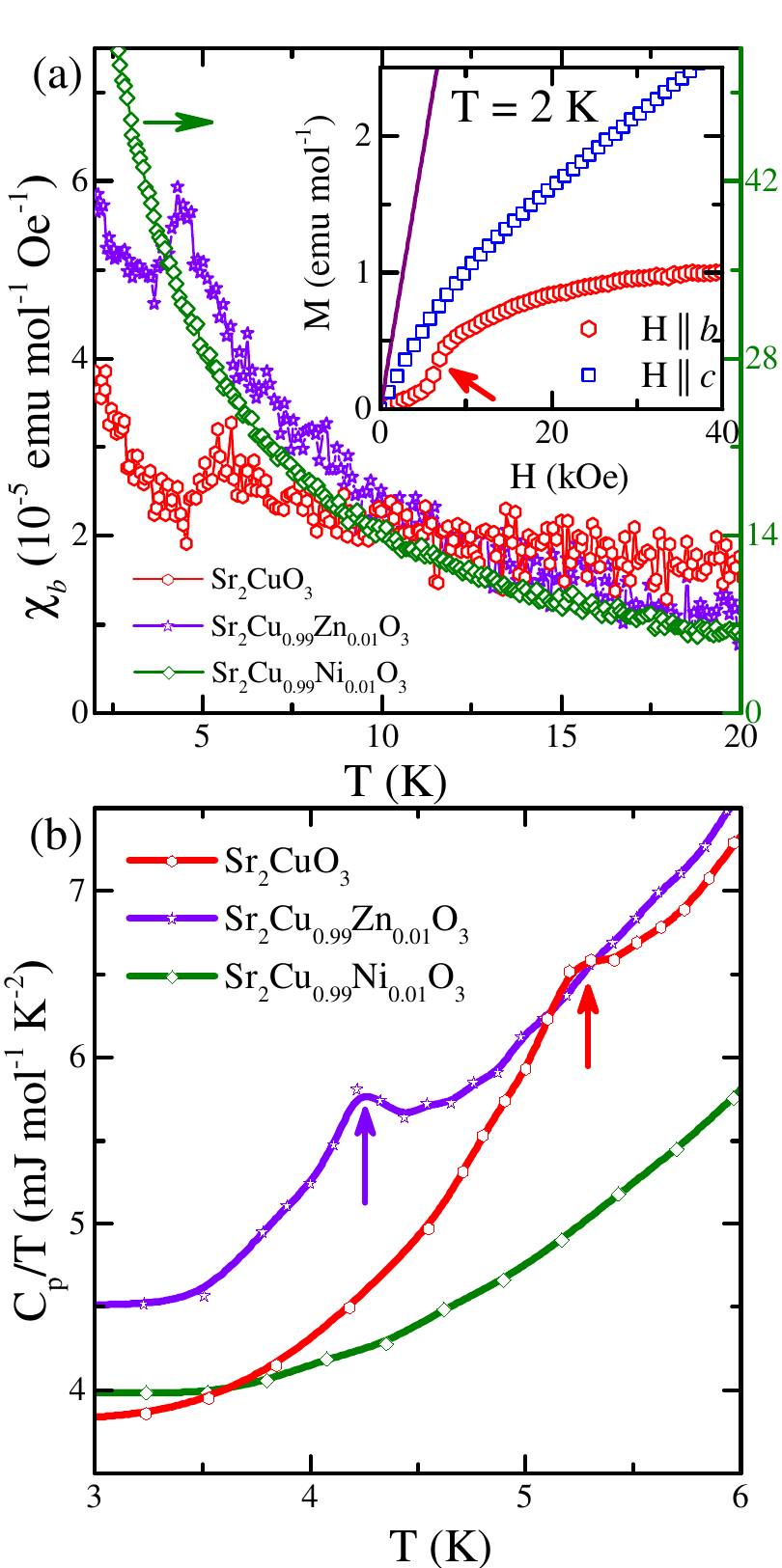}
\caption{(a) Low temperature susceptibilities $\chi$ of the pristine, Zn and Ni doped Sr$_{2}$CuO$_{3}$ crystals. Measurements were carried out under an applied field H = 1 kOe along the chain direction. Inset: isothermal magnetization M(H) of the pristine crystal along crystrallographic \textit{b} and \textit{c} axes. The arrow indicates the field induced transition in the H $\parallel$ \textit{b} measurements. The solid line is the calculated M(H) at T = 2 K (see text for details). (b) Specific heat $C_{p}/T$ is shown for the three compounds. The lines are guides to the eye. Arrows show the magnetic transitions.}
\label{Mag_order}
\end{figure}

\textbf{\textit{Magnetic ordering in doped Sr$_{2}$CuO$_{3}$}:} Here we briefly discuss the influence of doping on the magnetic ordering temperature. Previously, the weak magnetic ordering in Sr$_{2}$CuO$_{3}$ at T = 5.4 K was reported based on neutron and $\mu$SR studies\cite{Kojima18}. From neutron studies the ordered Cu moment is reported to be $\lesssim 0.06 \mu_{B}$\cite{Kojima18}. A small but clearly discernible peak near $T \approx 5.5$ K in our susceptibility data measured under an applied field of 1 kOe (Fig. \ref{Mag_order}) indicates a transition to an ordered state. The antiferromagnetic ordering of the Cu$^{2+}$ moments is also reflected in the isothermal magnetization of the pristine sample at T = 2 K for H $\parallel$ \textit{b} and H $\parallel$ \textit{c} (i.e., parallel to the chains and perpendicular to CuO$_{2}$ plaquettes, respectively). The data are shown as an inset in the panel (a) of Fig. \ref{Mag_order}. In the ordered ground state, M(H) exhibits a sharp field induced transition near H = 5 kOe along the chain direction (H $\parallel$ \textit{b}), while along H $\parallel$ \textit{c} it shows only a monotonic increase. The calculated M(H) for N$_{s}$ ~ 1.8 $\times 10^{-3}$ per Cu site, obtained using the ‘p’ value listed in table. \ref{FP}, is compared with the experimental data. With increasing field values, the calculated curve increases far more rapidly than the experimental data. This was expected, since at T = 2 K the spins are in an antiferromagnetically ordered state. Upon doping Zn the transition temperature is suppressed to $T = 4$ K. To confirm the bulk nature of these transitions, specific heat measurements were performed where anomalies were seen at the same temperatures as the susceptibilities (shown in  the inset of Fig. \ref{Mag_order}) where molar specific heat divided by temperature ($C_{P}/T$) is shown as a function of T. The detection of weak magnetic ordering in our bulk measurements is an indication of the high quality of the single crystals used in the present work. The transition temperature upon Pd doping in Sr$_{2}$CuO$_{3}$ is shown to decrease\cite{Kojima18} in agreement with a quantum mechanical calculation\cite{Eggert29}. The observed suppression of the transition temperature upon Zn doping is in a good qualitative agreement with this theory. The effect of Ni doping on the magnetic ordering temperature, however, is quite dramatic as no signs of transition to an ordered state either in the susceptibility or the specific heat are observed down to the lowest measurement temperature of T = 2 K in our experiments. According to the theory of ref. \cite{Eggert29}, the transition temperature under 1 \% of scalar impurity, should have been suppressed to about 0.5T$_N$ (see, Fig. 7 of ref. \cite{Kojima18}). However, the absence of any sign of ordering in our Ni-doped crystal probably indicates that  opening of a sizable spin gap has ousted the magnetic ordering completely or has suppressed it to temperatures far below T = 2 K. It should be noticed that the specific heat of Ni doped crystal at low temperatures is slightly (about $2-3 mJ/mol K^{2}$) smaller in magnitude compared to that of the pristine or Zn doped crystal. This decrease is probably a manifestation of the spin \textit{pseudogap}. A complete analysis of the specific heat in the absence of a lattice template and without the knowledge of contribution arising from the free spins at the chain ends is difficult at this stage. 

\section{Conclusion}
In conclusion, we studied the finite-size effects in the quasi-one-dimensional spin $S = \frac{1}{2}$ Heisenberg antiferromagnets Sr$_{2}$CuO$_{3}$, Sr$_{2}$Cu$_{0.99}$M$_{0.01}$O$_{3}$ (M = Zn and Ni) and SrCuO$_{2}$. The investigations were carried out on high-quality single-crystalline samples. We experimentally verified the field theory results published by Sirker \textit{et al.} \cite{Sirker09,Sirker12} describing the finite-size effects on the magnetic susceptibility of a spin $\frac{1}{2}$ Heisenberg Antiferromagnetic chain.We also showed that dilute concentration of Ni$^{2+}$ions in the chain behave like ‘scalar’ defects as predicted by the theory of ref. \cite{Eggert10}, which is probably related to gapping of the spinon excitation spectrum upon Ni doping as shown by Simutis \textit{et al.} \cite{Simutis30}. We also show the inadequacy of the model in describing the susceptibility in two cases: (i) when the ground state is gapped and (ii) in the zigzag chains characterized by the presence of weak rung coupling.\cite{Simutis30} Our work should motivate further field theoretic studies to extend the model in order to remove these inadequacies. In future, experimental verification of this theory on other spin chain compounds should be done.  The influence of doping on the low-temperature long-range spin ordering temperature was studied. Zn-doping suppresses the magnetic ordering in accordance with the theory. On the other hand, in the Ni doped crystal no signs of magnetic ordering were found down to the lowest measurement temperature of T = 2 K. The complete ousting of the magnetic ordering or its suppression far below T = 2 K is probably due to the gapped ground state. In the compound SrCuO$_{2}$, consisting of zigzag $S = \frac{1}{2}$ chains, the influence of spin frustration on the magnetic ordering and the defect concentration determined from the susceptibility data is discussed.

\begin{acknowledgments}
We are thankful to Jesko Sirker for proof reading the manuscript and making constructive suggestions. We also thank Nandini Trivedi and Christian R\"uegg for useful discussions. We acknowledge financial support under Project No. : INT/SWISS/ISJRP/PEP/P-06/2012. 
\end{acknowledgments}

\bibliographystyle{apsrev4-1}
\bibliography{Bibliography}

\begin{thebibliography}{42}%
\makeatletter
\providecommand \@ifxundefined [1]{%
 \@ifx{#1\undefined}
}%
\providecommand \@ifnum [1]{%
 \ifnum #1\expandafter \@firstoftwo
 \else \expandafter \@secondoftwo
 \fi
}%
\providecommand \@ifx [1]{%
 \ifx #1\expandafter \@firstoftwo
 \else \expandafter \@secondoftwo
 \fi
}%
\providecommand \natexlab [1]{#1}%
\providecommand \enquote  [1]{``#1''}%
\providecommand \bibnamefont  [1]{#1}%
\providecommand \bibfnamefont [1]{#1}%
\providecommand \citenamefont [1]{#1}%
\providecommand \href@noop [0]{\@secondoftwo}%
\providecommand \href [0]{\begingroup \@sanitize@url \@href}%
\providecommand \@href[1]{\@@startlink{#1}\@@href}%
\providecommand \@@href[1]{\endgroup#1\@@endlink}%
\providecommand \@sanitize@url [0]{\catcode `\\12\catcode `\$12\catcode
  `\&12\catcode `\#12\catcode `\^12\catcode `\_12\catcode `\%12\relax}%
\providecommand \@@startlink[1]{}%
\providecommand \@@endlink[0]{}%
\providecommand \url  [0]{\begingroup\@sanitize@url \@url }%
\providecommand \@url [1]{\endgroup\@href {#1}{\urlprefix }}%
\providecommand \urlprefix  [0]{URL }%
\providecommand \Eprint [0]{\href }%
\providecommand \doibase [0]{http://dx.doi.org/}%
\providecommand \selectlanguage [0]{\@gobble}%
\providecommand \bibinfo  [0]{\@secondoftwo}%
\providecommand \bibfield  [0]{\@secondoftwo}%
\providecommand \translation [1]{[#1]}%
\providecommand \BibitemOpen [0]{}%
\providecommand \bibitemStop [0]{}%
\providecommand \bibitemNoStop [0]{.\EOS\space}%
\providecommand \EOS [0]{\spacefactor3000\relax}%
\providecommand \BibitemShut  [1]{\csname bibitem#1\endcsname}%
\let\auto@bib@innerbib\@empty
%</preamble>
\bibitem [{\citenamefont {Bethe}(1931)}]{Bethe01}%
  \BibitemOpen
  \bibfield  {author} {\bibinfo {author} {\bibfnamefont {H.}~\bibnamefont
  {Bethe}},\ }\href {\doibase 10.1007/BF01341708} {\bibfield  {journal}
  {\bibinfo  {journal} {Zeitschrift für Physik}\ }\textbf {\bibinfo {volume}
  {71}},\ \bibinfo {pages} {205} (\bibinfo {year} {1931})}\BibitemShut
  {NoStop}%
\bibitem [{\citenamefont {Bonner}\ and\ \citenamefont
  {Fisher}(1964)}]{Bonner02}%
  \BibitemOpen
  \bibfield  {author} {\bibinfo {author} {\bibfnamefont {J.~C.}\ \bibnamefont
  {Bonner}}\ and\ \bibinfo {author} {\bibfnamefont {M.~E.}\ \bibnamefont
  {Fisher}},\ }\href {\doibase 10.1103/PhysRev.135.A640} {\bibfield  {journal}
  {\bibinfo  {journal} {Phys. Rev.}\ }\textbf {\bibinfo {volume} {135}},\
  \bibinfo {pages} {A640} (\bibinfo {year} {1964})}\BibitemShut {NoStop}%
\bibitem [{\citenamefont {Balents}(2010)}]{Balents03}%
  \BibitemOpen
  \bibfield  {author} {\bibinfo {author} {\bibfnamefont {L.}~\bibnamefont
  {Balents}},\ }\href {\doibase 10.1038/nature08917} {\bibfield  {journal}
  {\bibinfo  {journal} {Nature}\ }\textbf {\bibinfo {volume} {464}},\ \bibinfo
  {pages} {199} (\bibinfo {year} {2010})}\BibitemShut {NoStop}%
\bibitem [{\citenamefont {Griffiths}(1964)}]{Griffiths04}%
  \BibitemOpen
  \bibfield  {author} {\bibinfo {author} {\bibfnamefont {R.~B.}\ \bibnamefont
  {Griffiths}},\ }\href {\doibase 10.1103/PhysRev.133.A768} {\bibfield
  {journal} {\bibinfo  {journal} {Phys. Rev.}\ }\textbf {\bibinfo {volume}
  {133}},\ \bibinfo {pages} {A768} (\bibinfo {year} {1964})}\BibitemShut
  {NoStop}%
\bibitem [{\citenamefont {Eggert}\ \emph {et~al.}(1994)\citenamefont {Eggert},
  \citenamefont {Affleck},\ and\ \citenamefont {Takahashi}}]{Eggert22}%
  \BibitemOpen
  \bibfield  {author} {\bibinfo {author} {\bibfnamefont {S.}~\bibnamefont
  {Eggert}}, \bibinfo {author} {\bibfnamefont {I.}~\bibnamefont {Affleck}}, \
  and\ \bibinfo {author} {\bibfnamefont {M.}~\bibnamefont {Takahashi}},\ }\href
  {\doibase 10.1103/PhysRevLett.73.332} {\bibfield  {journal} {\bibinfo
  {journal} {Phys. Rev. Lett.}\ }\textbf {\bibinfo {volume} {73}},\ \bibinfo
  {pages} {332} (\bibinfo {year} {1994})}\BibitemShut {NoStop}%
\bibitem [{\citenamefont {Lukyanov}(1998)}]{Lukyanov06}%
  \BibitemOpen
  \bibfield  {author} {\bibinfo {author} {\bibfnamefont {S.}~\bibnamefont
  {Lukyanov}},\ }\href {\doibase
  http://dx.doi.org/10.1016/S0550-3213(98)00249-1} {\bibfield  {journal}
  {\bibinfo  {journal} {Nuclear Physics B}\ }\textbf {\bibinfo {volume}
  {522}},\ \bibinfo {pages} {533 } (\bibinfo {year} {1998})}\BibitemShut
  {NoStop}%
\bibitem [{\citenamefont {Kl\"umper}(1998)}]{Klumper05}%
  \BibitemOpen
  \bibfield  {author} {\bibinfo {author} {\bibfnamefont {A.}~\bibnamefont
  {Kl\"umper}},\ }\href {\doibase 10.1007/s100510050491} {\bibfield  {journal}
  {\bibinfo  {journal} {Eur. Phys. J. B}\ }\textbf {\bibinfo {volume} {5}},\
  \bibinfo {pages} {677} (\bibinfo {year} {1998})}\BibitemShut {NoStop}%
\bibitem [{\citenamefont {Kl\"umper}\ and\ \citenamefont
  {Johnston}(2000)}]{Klumper33}%
  \BibitemOpen
  \bibfield  {author} {\bibinfo {author} {\bibfnamefont {A.}~\bibnamefont
  {Kl\"umper}}\ and\ \bibinfo {author} {\bibfnamefont {D.~C.}\ \bibnamefont
  {Johnston}},\ }\href {\doibase 10.1103/PhysRevLett.84.4701} {\bibfield
  {journal} {\bibinfo  {journal} {Phys. Rev. Lett.}\ }\textbf {\bibinfo
  {volume} {84}},\ \bibinfo {pages} {4701} (\bibinfo {year}
  {2000})}\BibitemShut {NoStop}%
\bibitem [{\citenamefont {Janson}\ \emph {et~al.}(2009)\citenamefont {Janson},
  \citenamefont {Schnelle}, \citenamefont {Schmidt}, \citenamefont {Prots},
  \citenamefont {Drechsler}, \citenamefont {Filatov},\ and\ \citenamefont
  {Rosner}}]{Janson34}%
  \BibitemOpen
  \bibfield  {author} {\bibinfo {author} {\bibfnamefont {O.}~\bibnamefont
  {Janson}}, \bibinfo {author} {\bibfnamefont {W.}~\bibnamefont {Schnelle}},
  \bibinfo {author} {\bibfnamefont {M.}~\bibnamefont {Schmidt}}, \bibinfo
  {author} {\bibfnamefont {Y.}~\bibnamefont {Prots}}, \bibinfo {author}
  {\bibfnamefont {S.-L.}\ \bibnamefont {Drechsler}}, \bibinfo {author}
  {\bibfnamefont {S.~K.}\ \bibnamefont {Filatov}}, \ and\ \bibinfo {author}
  {\bibfnamefont {H.}~\bibnamefont {Rosner}},\ }\href
  {http://stacks.iop.org/1367-2630/11/i=11/a=113034} {\bibfield  {journal}
  {\bibinfo  {journal} {New Journal of Physics}\ }\textbf {\bibinfo {volume}
  {11}},\ \bibinfo {pages} {113034} (\bibinfo {year} {2009})}\BibitemShut
  {NoStop}%
\bibitem [{\citenamefont {Fujimoto}\ and\ \citenamefont
  {Eggert}(2004)}]{Fujimoto19}%
  \BibitemOpen
  \bibfield  {author} {\bibinfo {author} {\bibfnamefont {S.}~\bibnamefont
  {Fujimoto}}\ and\ \bibinfo {author} {\bibfnamefont {S.}~\bibnamefont
  {Eggert}},\ }\href {\doibase 10.1103/PhysRevLett.92.037206} {\bibfield
  {journal} {\bibinfo  {journal} {Phys. Rev. Lett.}\ }\textbf {\bibinfo
  {volume} {92}},\ \bibinfo {pages} {037206} (\bibinfo {year}
  {2004})}\BibitemShut {NoStop}%
\bibitem [{\citenamefont {Sirker}\ \emph {et~al.}(2007)\citenamefont {Sirker},
  \citenamefont {Laflorencie}, \citenamefont {Fujimoto}, \citenamefont
  {Eggert},\ and\ \citenamefont {Affleck}}]{Sirker09}%
  \BibitemOpen
  \bibfield  {author} {\bibinfo {author} {\bibfnamefont {J.}~\bibnamefont
  {Sirker}}, \bibinfo {author} {\bibfnamefont {N.}~\bibnamefont {Laflorencie}},
  \bibinfo {author} {\bibfnamefont {S.}~\bibnamefont {Fujimoto}}, \bibinfo
  {author} {\bibfnamefont {S.}~\bibnamefont {Eggert}}, \ and\ \bibinfo {author}
  {\bibfnamefont {I.}~\bibnamefont {Affleck}},\ }\href {\doibase
  10.1103/PhysRevLett.98.137205} {\bibfield  {journal} {\bibinfo  {journal}
  {Phys. Rev. Lett.}\ }\textbf {\bibinfo {volume} {98}},\ \bibinfo {pages}
  {137205} (\bibinfo {year} {2007})}\BibitemShut {NoStop}%
\bibitem [{\citenamefont {Ami}\ \emph {et~al.}(1995)\citenamefont {Ami},
  \citenamefont {Crawford}, \citenamefont {Harlow}, \citenamefont {Wang},
  \citenamefont {Johnston}, \citenamefont {Huang},\ and\ \citenamefont
  {Erwin}}]{Ami07}%
  \BibitemOpen
  \bibfield  {author} {\bibinfo {author} {\bibfnamefont {T.}~\bibnamefont
  {Ami}}, \bibinfo {author} {\bibfnamefont {M.~K.}\ \bibnamefont {Crawford}},
  \bibinfo {author} {\bibfnamefont {R.~L.}\ \bibnamefont {Harlow}}, \bibinfo
  {author} {\bibfnamefont {Z.~R.}\ \bibnamefont {Wang}}, \bibinfo {author}
  {\bibfnamefont {D.~C.}\ \bibnamefont {Johnston}}, \bibinfo {author}
  {\bibfnamefont {Q.}~\bibnamefont {Huang}}, \ and\ \bibinfo {author}
  {\bibfnamefont {R.~W.}\ \bibnamefont {Erwin}},\ }\href {\doibase
  10.1103/PhysRevB.51.5994} {\bibfield  {journal} {\bibinfo  {journal} {Phys.
  Rev. B}\ }\textbf {\bibinfo {volume} {51}},\ \bibinfo {pages} {5994}
  (\bibinfo {year} {1995})}\BibitemShut {NoStop}%
\bibitem [{\citenamefont {Chattopadhyay}\ \emph {et~al.}(2011)\citenamefont
  {Chattopadhyay}, \citenamefont {Giri},\ and\ \citenamefont
  {Majumdar}}]{Chattopadhyay35}%
  \BibitemOpen
  \bibfield  {author} {\bibinfo {author} {\bibfnamefont {S.}~\bibnamefont
  {Chattopadhyay}}, \bibinfo {author} {\bibfnamefont {S.}~\bibnamefont {Giri}},
  \ and\ \bibinfo {author} {\bibfnamefont {S.}~\bibnamefont {Majumdar}},\
  }\href {http://stacks.iop.org/0953-8984/23/i=21/a=216006} {\bibfield
  {journal} {\bibinfo  {journal} {Journal of Physics: Condensed Matter}\
  }\textbf {\bibinfo {volume} {23}},\ \bibinfo {pages} {216006} (\bibinfo
  {year} {2011})}\BibitemShut {NoStop}%
\bibitem [{\citenamefont {Motoyama}\ \emph {et~al.}(1996)\citenamefont
  {Motoyama}, \citenamefont {Eisaki},\ and\ \citenamefont
  {Uchida}}]{Motoyama08}%
  \BibitemOpen
  \bibfield  {author} {\bibinfo {author} {\bibfnamefont {N.}~\bibnamefont
  {Motoyama}}, \bibinfo {author} {\bibfnamefont {H.}~\bibnamefont {Eisaki}}, \
  and\ \bibinfo {author} {\bibfnamefont {S.}~\bibnamefont {Uchida}},\ }\href
  {\doibase 10.1103/PhysRevLett.76.3212} {\bibfield  {journal} {\bibinfo
  {journal} {Phys. Rev. Lett.}\ }\textbf {\bibinfo {volume} {76}},\ \bibinfo
  {pages} {3212} (\bibinfo {year} {1996})}\BibitemShut {NoStop}%
\bibitem [{\citenamefont {Eggert}\ and\ \citenamefont
  {Affleck}(1992)}]{Eggert10}%
  \BibitemOpen
  \bibfield  {author} {\bibinfo {author} {\bibfnamefont {S.}~\bibnamefont
  {Eggert}}\ and\ \bibinfo {author} {\bibfnamefont {I.}~\bibnamefont
  {Affleck}},\ }\href {\doibase 10.1103/PhysRevB.46.10866} {\bibfield
  {journal} {\bibinfo  {journal} {Phys. Rev. B}\ }\textbf {\bibinfo {volume}
  {46}},\ \bibinfo {pages} {10866} (\bibinfo {year} {1992})}\BibitemShut
  {NoStop}%
\bibitem [{\citenamefont {Karmakar}\ \emph {et~al.}(2014)\citenamefont
  {Karmakar}, \citenamefont {Singh}, \citenamefont {Singh}, \citenamefont
  {Poole},\ and\ \citenamefont {R\"uegg}}]{Karmakar13}%
  \BibitemOpen
  \bibfield  {author} {\bibinfo {author} {\bibfnamefont {K.}~\bibnamefont
  {Karmakar}}, \bibinfo {author} {\bibfnamefont {A.}~\bibnamefont {Singh}},
  \bibinfo {author} {\bibfnamefont {S.}~\bibnamefont {Singh}}, \bibinfo
  {author} {\bibfnamefont {A.}~\bibnamefont {Poole}}, \ and\ \bibinfo {author}
  {\bibfnamefont {C.}~\bibnamefont {R\"uegg}},\ }\href {\doibase
  10.1021/cg4016993} {\bibfield  {journal} {\bibinfo  {journal} {Crystal Growth
  \& Design}\ }\textbf {\bibinfo {volume} {14}},\ \bibinfo {pages} {1184}
  (\bibinfo {year} {2014})}\BibitemShut {NoStop}%
\bibitem [{\citenamefont {Pfann}(1966)}]{Pfann14}%
  \BibitemOpen
  \bibfield  {author} {\bibinfo {author} {\bibfnamefont {W.}~\bibnamefont
  {Pfann}},\ }\href@noop {} {\emph {\bibinfo {title} {Zone Melting}}}\
  (\bibinfo  {publisher} {Wiley \& Sons, New York},\ \bibinfo {year}
  {1966})\BibitemShut {NoStop}%
\bibitem [{\citenamefont {Dhalenne}\ \emph {et~al.}(1997)\citenamefont
  {Dhalenne}, \citenamefont {Revcolevschi}, \citenamefont {Rouchaud},\ and\
  \citenamefont {Federoff}}]{Revcolevschi36}%
  \BibitemOpen
  \bibfield  {author} {\bibinfo {author} {\bibfnamefont {G.}~\bibnamefont
  {Dhalenne}}, \bibinfo {author} {\bibfnamefont {A.}~\bibnamefont
  {Revcolevschi}}, \bibinfo {author} {\bibfnamefont {J.}~\bibnamefont
  {Rouchaud}}, \ and\ \bibinfo {author} {\bibfnamefont {M.}~\bibnamefont
  {Federoff}},\ }\href {\doibase
  http://dx.doi.org/10.1016/S0025-5408(97)00058-5} {\bibfield  {journal}
  {\bibinfo  {journal} {Materials Research Bulletin}\ }\textbf {\bibinfo
  {volume} {32}},\ \bibinfo {pages} {939 } (\bibinfo {year}
  {1997})}\BibitemShut {NoStop}%
\bibitem [{\citenamefont {Rice}\ \emph {et~al.}(1993)\citenamefont {Rice},
  \citenamefont {Gopalan},\ and\ \citenamefont {Sigrist}}]{Rice15}%
  \BibitemOpen
  \bibfield  {author} {\bibinfo {author} {\bibfnamefont {T.~M.}\ \bibnamefont
  {Rice}}, \bibinfo {author} {\bibfnamefont {S.}~\bibnamefont {Gopalan}}, \
  and\ \bibinfo {author} {\bibfnamefont {M.}~\bibnamefont {Sigrist}},\ }\href
  {http://stacks.iop.org/0295-5075/23/i=6/a=011} {\bibfield  {journal}
  {\bibinfo  {journal} {EPL (Europhysics Letters)}\ }\textbf {\bibinfo {volume}
  {23}},\ \bibinfo {pages} {445} (\bibinfo {year} {1993})}\BibitemShut
  {NoStop}%
\bibitem [{\citenamefont {Zaliznyak}\ \emph {et~al.}(2004)\citenamefont
  {Zaliznyak}, \citenamefont {Woo}, \citenamefont {Perring}, \citenamefont
  {Broholm}, \citenamefont {Frost},\ and\ \citenamefont
  {Takagi}}]{Zaliznyak16}%
  \BibitemOpen
  \bibfield  {author} {\bibinfo {author} {\bibfnamefont {I.~A.}\ \bibnamefont
  {Zaliznyak}}, \bibinfo {author} {\bibfnamefont {H.}~\bibnamefont {Woo}},
  \bibinfo {author} {\bibfnamefont {T.~G.}\ \bibnamefont {Perring}}, \bibinfo
  {author} {\bibfnamefont {C.~L.}\ \bibnamefont {Broholm}}, \bibinfo {author}
  {\bibfnamefont {C.~D.}\ \bibnamefont {Frost}}, \ and\ \bibinfo {author}
  {\bibfnamefont {H.}~\bibnamefont {Takagi}},\ }\href {\doibase
  10.1103/PhysRevLett.93.087202} {\bibfield  {journal} {\bibinfo  {journal}
  {Phys. Rev. Lett.}\ }\textbf {\bibinfo {volume} {93}},\ \bibinfo {pages}
  {087202} (\bibinfo {year} {2004})}\BibitemShut {NoStop}%
\bibitem [{\citenamefont {Keren}\ \emph {et~al.}(1993)\citenamefont {Keren},
  \citenamefont {Le}, \citenamefont {Luke}, \citenamefont {Sternlieb},
  \citenamefont {Wu}, \citenamefont {Uemura}, \citenamefont {Tajima},\ and\
  \citenamefont {Uchida}}]{Keren17}%
  \BibitemOpen
  \bibfield  {author} {\bibinfo {author} {\bibfnamefont {A.}~\bibnamefont
  {Keren}}, \bibinfo {author} {\bibfnamefont {L.~P.}\ \bibnamefont {Le}},
  \bibinfo {author} {\bibfnamefont {G.~M.}\ \bibnamefont {Luke}}, \bibinfo
  {author} {\bibfnamefont {B.~J.}\ \bibnamefont {Sternlieb}}, \bibinfo {author}
  {\bibfnamefont {W.~D.}\ \bibnamefont {Wu}}, \bibinfo {author} {\bibfnamefont
  {Y.~J.}\ \bibnamefont {Uemura}}, \bibinfo {author} {\bibfnamefont
  {S.}~\bibnamefont {Tajima}}, \ and\ \bibinfo {author} {\bibfnamefont
  {S.}~\bibnamefont {Uchida}},\ }\href {\doibase 10.1103/PhysRevB.48.12926}
  {\bibfield  {journal} {\bibinfo  {journal} {Phys. Rev. B}\ }\textbf {\bibinfo
  {volume} {48}},\ \bibinfo {pages} {12926} (\bibinfo {year}
  {1993})}\BibitemShut {NoStop}%
\bibitem [{\citenamefont {Matsuda}\ \emph {et~al.}(1997)\citenamefont
  {Matsuda}, \citenamefont {Katsumata}, \citenamefont {Kojima}, \citenamefont
  {Larkin}, \citenamefont {Luke}, \citenamefont {Merrin}, \citenamefont
  {Nachumi}, \citenamefont {Uemura}, \citenamefont {Eisaki}, \citenamefont
  {Motoyama}, \citenamefont {Uchida},\ and\ \citenamefont
  {Shirane}}]{Matsuda32}%
  \BibitemOpen
  \bibfield  {author} {\bibinfo {author} {\bibfnamefont {M.}~\bibnamefont
  {Matsuda}}, \bibinfo {author} {\bibfnamefont {K.}~\bibnamefont {Katsumata}},
  \bibinfo {author} {\bibfnamefont {K.~M.}\ \bibnamefont {Kojima}}, \bibinfo
  {author} {\bibfnamefont {M.}~\bibnamefont {Larkin}}, \bibinfo {author}
  {\bibfnamefont {G.~M.}\ \bibnamefont {Luke}}, \bibinfo {author}
  {\bibfnamefont {J.}~\bibnamefont {Merrin}}, \bibinfo {author} {\bibfnamefont
  {B.}~\bibnamefont {Nachumi}}, \bibinfo {author} {\bibfnamefont {Y.~J.}\
  \bibnamefont {Uemura}}, \bibinfo {author} {\bibfnamefont {H.}~\bibnamefont
  {Eisaki}}, \bibinfo {author} {\bibfnamefont {N.}~\bibnamefont {Motoyama}},
  \bibinfo {author} {\bibfnamefont {S.}~\bibnamefont {Uchida}}, \ and\ \bibinfo
  {author} {\bibfnamefont {G.}~\bibnamefont {Shirane}},\ }\href {\doibase
  10.1103/PhysRevB.55.R11953} {\bibfield  {journal} {\bibinfo  {journal} {Phys.
  Rev. B}\ }\textbf {\bibinfo {volume} {55}},\ \bibinfo {pages} {R11953}
  (\bibinfo {year} {1997})}\BibitemShut {NoStop}%
\bibitem [{\citenamefont {Kojima}\ \emph {et~al.}(1997)\citenamefont {Kojima},
  \citenamefont {Fudamoto}, \citenamefont {Larkin}, \citenamefont {Luke},
  \citenamefont {Merrin}, \citenamefont {Nachumi}, \citenamefont {Uemura},
  \citenamefont {Motoyama}, \citenamefont {Eisaki}, \citenamefont {Uchida},
  \citenamefont {Yamada}, \citenamefont {Endoh}, \citenamefont {Hosoya},
  \citenamefont {Sternlieb},\ and\ \citenamefont {Shirane}}]{Kojima18}%
  \BibitemOpen
  \bibfield  {author} {\bibinfo {author} {\bibfnamefont {K.~M.}\ \bibnamefont
  {Kojima}}, \bibinfo {author} {\bibfnamefont {Y.}~\bibnamefont {Fudamoto}},
  \bibinfo {author} {\bibfnamefont {M.}~\bibnamefont {Larkin}}, \bibinfo
  {author} {\bibfnamefont {G.~M.}\ \bibnamefont {Luke}}, \bibinfo {author}
  {\bibfnamefont {J.}~\bibnamefont {Merrin}}, \bibinfo {author} {\bibfnamefont
  {B.}~\bibnamefont {Nachumi}}, \bibinfo {author} {\bibfnamefont {Y.~J.}\
  \bibnamefont {Uemura}}, \bibinfo {author} {\bibfnamefont {N.}~\bibnamefont
  {Motoyama}}, \bibinfo {author} {\bibfnamefont {H.}~\bibnamefont {Eisaki}},
  \bibinfo {author} {\bibfnamefont {S.}~\bibnamefont {Uchida}}, \bibinfo
  {author} {\bibfnamefont {K.}~\bibnamefont {Yamada}}, \bibinfo {author}
  {\bibfnamefont {Y.}~\bibnamefont {Endoh}}, \bibinfo {author} {\bibfnamefont
  {S.}~\bibnamefont {Hosoya}}, \bibinfo {author} {\bibfnamefont {B.~J.}\
  \bibnamefont {Sternlieb}}, \ and\ \bibinfo {author} {\bibfnamefont
  {G.}~\bibnamefont {Shirane}},\ }\href {\doibase 10.1103/PhysRevLett.78.1787}
  {\bibfield  {journal} {\bibinfo  {journal} {Phys. Rev. Lett.}\ }\textbf
  {\bibinfo {volume} {78}},\ \bibinfo {pages} {1787} (\bibinfo {year}
  {1997})}\BibitemShut {NoStop}%
\bibitem [{\citenamefont {Sirker}\ \emph {et~al.}(2008)\citenamefont {Sirker},
  \citenamefont {Fujimoto}, \citenamefont {Laflorencie}, \citenamefont
  {Eggert},\ and\ \citenamefont {Affleck}}]{Sirker12}%
  \BibitemOpen
  \bibfield  {author} {\bibinfo {author} {\bibfnamefont {J.}~\bibnamefont
  {Sirker}}, \bibinfo {author} {\bibfnamefont {S.}~\bibnamefont {Fujimoto}},
  \bibinfo {author} {\bibfnamefont {N.}~\bibnamefont {Laflorencie}}, \bibinfo
  {author} {\bibfnamefont {S.}~\bibnamefont {Eggert}}, \ and\ \bibinfo {author}
  {\bibfnamefont {I.}~\bibnamefont {Affleck}},\ }\href
  {http://stacks.iop.org/1742-5468/2008/i=02/a=P02015} {\bibfield  {journal}
  {\bibinfo  {journal} {Journal of Statistical Mechanics: Theory and
  Experiment}\ }\textbf {\bibinfo {volume} {2008}},\ \bibinfo {pages} {P02015}
  (\bibinfo {year} {2008})}\BibitemShut {NoStop}%
\bibitem [{\citenamefont {Mahajan}\ and\ \citenamefont
  {Venkataramani}(2001)}]{Mahajan41}%
  \BibitemOpen
  \bibfield  {author} {\bibinfo {author} {\bibfnamefont {A.~V.}\ \bibnamefont
  {Mahajan}}\ and\ \bibinfo {author} {\bibfnamefont {N.}~\bibnamefont
  {Venkataramani}},\ }\href {\doibase 10.1103/PhysRevB.64.092410} {\bibfield
  {journal} {\bibinfo  {journal} {Phys. Rev. B}\ }\textbf {\bibinfo {volume}
  {64}},\ \bibinfo {pages} {092410} (\bibinfo {year} {2001})}\BibitemShut
  {NoStop}%
\bibitem [{\citenamefont {Ohta}\ \emph {et~al.}(1992)\citenamefont {Ohta},
  \citenamefont {Yamauchi}, \citenamefont {Motokawa}, \citenamefont {Azuma},\
  and\ \citenamefont {Takano}}]{Ohta42}%
  \BibitemOpen
  \bibfield  {author} {\bibinfo {author} {\bibfnamefont {H.}~\bibnamefont
  {Ohta}}, \bibinfo {author} {\bibfnamefont {N.}~\bibnamefont {Yamauchi}},
  \bibinfo {author} {\bibfnamefont {M.}~\bibnamefont {Motokawa}}, \bibinfo
  {author} {\bibfnamefont {M.}~\bibnamefont {Azuma}}, \ and\ \bibinfo {author}
  {\bibfnamefont {M.}~\bibnamefont {Takano}},\ }\href
  {http://journals.jps.jp/doi/abs/10.1143/JPSJ.61.3370} {\bibfield  {journal}
  {\bibinfo  {journal} {Journal of Physical Society of Japan}\ }\textbf
  {\bibinfo {volume} {61}},\ \bibinfo {pages} {3370} (\bibinfo {year}
  {1992})}\BibitemShut {NoStop}%
\bibitem [{\citenamefont {Asakawa}\ \emph {et~al.}(1998)\citenamefont
  {Asakawa}, \citenamefont {Matsuda}, \citenamefont {Minami}, \citenamefont
  {Yamazaki},\ and\ \citenamefont {Katsumata}}]{Asakawa44}%
  \BibitemOpen
  \bibfield  {author} {\bibinfo {author} {\bibfnamefont {H.}~\bibnamefont
  {Asakawa}}, \bibinfo {author} {\bibfnamefont {M.}~\bibnamefont {Matsuda}},
  \bibinfo {author} {\bibfnamefont {K.}~\bibnamefont {Minami}}, \bibinfo
  {author} {\bibfnamefont {H.}~\bibnamefont {Yamazaki}}, \ and\ \bibinfo
  {author} {\bibfnamefont {K.}~\bibnamefont {Katsumata}},\ }\href {\doibase
  10.1103/PhysRevB.57.8285} {\bibfield  {journal} {\bibinfo  {journal} {Phys.
  Rev. B}\ }\textbf {\bibinfo {volume} {57}},\ \bibinfo {pages} {8285}
  (\bibinfo {year} {1998})}\BibitemShut {NoStop}%
\bibitem [{\citenamefont {Johnston}\ \emph {et~al.}(2000)\citenamefont
  {Johnston}, \citenamefont {Kremer}, \citenamefont {Troyer}, \citenamefont
  {Wang}, \citenamefont {Kl\"umper}, \citenamefont {Bud'ko}, \citenamefont
  {Panchula},\ and\ \citenamefont {Canfield}}]{Johnston21}%
  \BibitemOpen
  \bibfield  {author} {\bibinfo {author} {\bibfnamefont {D.~C.}\ \bibnamefont
  {Johnston}}, \bibinfo {author} {\bibfnamefont {R.~K.}\ \bibnamefont
  {Kremer}}, \bibinfo {author} {\bibfnamefont {M.}~\bibnamefont {Troyer}},
  \bibinfo {author} {\bibfnamefont {X.}~\bibnamefont {Wang}}, \bibinfo {author}
  {\bibfnamefont {A.}~\bibnamefont {Kl\"umper}}, \bibinfo {author}
  {\bibfnamefont {S.~L.}\ \bibnamefont {Bud'ko}}, \bibinfo {author}
  {\bibfnamefont {A.~F.}\ \bibnamefont {Panchula}}, \ and\ \bibinfo {author}
  {\bibfnamefont {P.~C.}\ \bibnamefont {Canfield}},\ }\href {\doibase
  10.1103/PhysRevB.61.9558} {\bibfield  {journal} {\bibinfo  {journal} {Phys.
  Rev. B}\ }\textbf {\bibinfo {volume} {61}},\ \bibinfo {pages} {9558}
  (\bibinfo {year} {2000})}\BibitemShut {NoStop}%
\bibitem [{\citenamefont {Suzuura}\ \emph {et~al.}(1996)\citenamefont
  {Suzuura}, \citenamefont {Yasuhara}, \citenamefont {Furusaki}, \citenamefont
  {Nagaosa},\ and\ \citenamefont {Tokura}}]{Suzuura20}%
  \BibitemOpen
  \bibfield  {author} {\bibinfo {author} {\bibfnamefont {H.}~\bibnamefont
  {Suzuura}}, \bibinfo {author} {\bibfnamefont {H.}~\bibnamefont {Yasuhara}},
  \bibinfo {author} {\bibfnamefont {A.}~\bibnamefont {Furusaki}}, \bibinfo
  {author} {\bibfnamefont {N.}~\bibnamefont {Nagaosa}}, \ and\ \bibinfo
  {author} {\bibfnamefont {Y.}~\bibnamefont {Tokura}},\ }\href {\doibase
  10.1103/PhysRevLett.76.2579} {\bibfield  {journal} {\bibinfo  {journal}
  {Phys. Rev. Lett.}\ }\textbf {\bibinfo {volume} {76}},\ \bibinfo {pages}
  {2579} (\bibinfo {year} {1996})}\BibitemShut {NoStop}%
\bibitem [{\citenamefont {Sologubenko}\ \emph {et~al.}(2000)\citenamefont
  {Sologubenko}, \citenamefont {Felder}, \citenamefont {Giann\`o},
  \citenamefont {Ott}, \citenamefont {Vietkine},\ and\ \citenamefont
  {Revcolevschi}}]{Sologubenko37}%
  \BibitemOpen
  \bibfield  {author} {\bibinfo {author} {\bibfnamefont {A.~V.}\ \bibnamefont
  {Sologubenko}}, \bibinfo {author} {\bibfnamefont {E.}~\bibnamefont {Felder}},
  \bibinfo {author} {\bibfnamefont {K.}~\bibnamefont {Giann\`o}}, \bibinfo
  {author} {\bibfnamefont {H.~R.}\ \bibnamefont {Ott}}, \bibinfo {author}
  {\bibfnamefont {A.}~\bibnamefont {Vietkine}}, \ and\ \bibinfo {author}
  {\bibfnamefont {A.}~\bibnamefont {Revcolevschi}},\ }\href {\doibase
  10.1103/PhysRevB.62.R6108} {\bibfield  {journal} {\bibinfo  {journal} {Phys.
  Rev. B}\ }\textbf {\bibinfo {volume} {62}},\ \bibinfo {pages} {R6108}
  (\bibinfo {year} {2000})}\BibitemShut {NoStop}%
\bibitem [{\citenamefont {Sandvik}\ \emph {et~al.}(1997)\citenamefont
  {Sandvik}, \citenamefont {Singh},\ and\ \citenamefont
  {Campbell}}]{Sandvik23}%
  \BibitemOpen
  \bibfield  {author} {\bibinfo {author} {\bibfnamefont {A.~W.}\ \bibnamefont
  {Sandvik}}, \bibinfo {author} {\bibfnamefont {R.~R.~P.}\ \bibnamefont
  {Singh}}, \ and\ \bibinfo {author} {\bibfnamefont {D.~K.}\ \bibnamefont
  {Campbell}},\ }\href {\doibase 10.1103/PhysRevB.56.14510} {\bibfield
  {journal} {\bibinfo  {journal} {Phys. Rev. B}\ }\textbf {\bibinfo {volume}
  {56}},\ \bibinfo {pages} {14510} (\bibinfo {year} {1997})}\BibitemShut
  {NoStop}%
\bibitem [{\citenamefont {Ashcroft}\ and\ \citenamefont
  {Mermin}(1976)}]{Ashcroft31}%
  \BibitemOpen
  \bibfield  {author} {\bibinfo {author} {\bibfnamefont {N.}~\bibnamefont
  {Ashcroft}}\ and\ \bibinfo {author} {\bibfnamefont {N.}~\bibnamefont
  {Mermin}},\ }\href {http://books.google.co.in/books?id=oXIfAQAAMAAJ} {\emph
  {\bibinfo {title} {Solid State Physics}}},\ HRW international editions\
  (\bibinfo  {publisher} {Holt, Rinehart and Winston},\ \bibinfo {year}
  {1976})\BibitemShut {NoStop}%
\bibitem [{\citenamefont {Takigawa}\ \emph {et~al.}(1997)\citenamefont
  {Takigawa}, \citenamefont {Motoyama}, \citenamefont {Eisaki},\ and\
  \citenamefont {Uchida}}]{Takigawa25}%
  \BibitemOpen
  \bibfield  {author} {\bibinfo {author} {\bibfnamefont {M.}~\bibnamefont
  {Takigawa}}, \bibinfo {author} {\bibfnamefont {N.}~\bibnamefont {Motoyama}},
  \bibinfo {author} {\bibfnamefont {H.}~\bibnamefont {Eisaki}}, \ and\ \bibinfo
  {author} {\bibfnamefont {S.}~\bibnamefont {Uchida}},\ }\href {\doibase
  10.1103/PhysRevB.55.14129} {\bibfield  {journal} {\bibinfo  {journal} {Phys.
  Rev. B}\ }\textbf {\bibinfo {volume} {55}},\ \bibinfo {pages} {14129}
  (\bibinfo {year} {1997})}\BibitemShut {NoStop}%
\bibitem [{\citenamefont {Eggert}\ and\ \citenamefont
  {Affleck}(1995)}]{Eggert11}%
  \BibitemOpen
  \bibfield  {author} {\bibinfo {author} {\bibfnamefont {S.}~\bibnamefont
  {Eggert}}\ and\ \bibinfo {author} {\bibfnamefont {I.}~\bibnamefont
  {Affleck}},\ }\href {\doibase 10.1103/PhysRevLett.75.934} {\bibfield
  {journal} {\bibinfo  {journal} {Phys. Rev. Lett.}\ }\textbf {\bibinfo
  {volume} {75}},\ \bibinfo {pages} {934} (\bibinfo {year} {1995})}\BibitemShut
  {NoStop}%
\bibitem [{\citenamefont {Boucher}\ and\ \citenamefont
  {Takigawa}(2000)}]{Boucher26}%
  \BibitemOpen
  \bibfield  {author} {\bibinfo {author} {\bibfnamefont {J.~P.}\ \bibnamefont
  {Boucher}}\ and\ \bibinfo {author} {\bibfnamefont {M.}~\bibnamefont
  {Takigawa}},\ }\href {\doibase 10.1103/PhysRevB.62.367} {\bibfield  {journal}
  {\bibinfo  {journal} {Phys. Rev. B}\ }\textbf {\bibinfo {volume} {62}},\
  \bibinfo {pages} {367} (\bibinfo {year} {2000})}\BibitemShut {NoStop}%
\bibitem [{\citenamefont {Sirker}\ and\ \citenamefont
  {Laflorencie}(2009)}]{Sirker27}%
  \BibitemOpen
  \bibfield  {author} {\bibinfo {author} {\bibfnamefont {J.}~\bibnamefont
  {Sirker}}\ and\ \bibinfo {author} {\bibfnamefont {N.}~\bibnamefont
  {Laflorencie}},\ }\href {http://stacks.iop.org/0295-5075/86/i=5/a=57004}
  {\bibfield  {journal} {\bibinfo  {journal} {EPL (Europhysics Letters)}\
  }\textbf {\bibinfo {volume} {86}},\ \bibinfo {pages} {57004} (\bibinfo {year}
  {2009})}\BibitemShut {NoStop}%
\bibitem [{\citenamefont {Rosner}\ \emph {et~al.}(1997)\citenamefont {Rosner},
  \citenamefont {Eschrig}, \citenamefont {Hayn}, \citenamefont {Drechsler},\
  and\ \citenamefont {M\'alek}}]{Rosner28}%
  \BibitemOpen
  \bibfield  {author} {\bibinfo {author} {\bibfnamefont {H.}~\bibnamefont
  {Rosner}}, \bibinfo {author} {\bibfnamefont {H.}~\bibnamefont {Eschrig}},
  \bibinfo {author} {\bibfnamefont {R.}~\bibnamefont {Hayn}}, \bibinfo {author}
  {\bibfnamefont {S.-L.}\ \bibnamefont {Drechsler}}, \ and\ \bibinfo {author}
  {\bibfnamefont {J.}~\bibnamefont {M\'alek}},\ }\href {\doibase
  10.1103/PhysRevB.56.3402} {\bibfield  {journal} {\bibinfo  {journal} {Phys.
  Rev. B}\ }\textbf {\bibinfo {volume} {56}},\ \bibinfo {pages} {3402}
  (\bibinfo {year} {1997})}\BibitemShut {NoStop}%
\bibitem [{\citenamefont {Simutis}\ \emph {et~al.}(2013)\citenamefont
  {Simutis}, \citenamefont {Gvasaliya}, \citenamefont {M\aa{}nsson},
  \citenamefont {Chernyshev}, \citenamefont {Mohan}, \citenamefont {Singh},
  \citenamefont {Hess}, \citenamefont {Savici}, \citenamefont {Kolesnikov},
  \citenamefont {Piovano}, \citenamefont {Perring}, \citenamefont {Zaliznyak},
  \citenamefont {B\"uchner},\ and\ \citenamefont {Zheludev}}]{Simutis30}%
  \BibitemOpen
  \bibfield  {author} {\bibinfo {author} {\bibfnamefont {G.}~\bibnamefont
  {Simutis}}, \bibinfo {author} {\bibfnamefont {S.}~\bibnamefont {Gvasaliya}},
  \bibinfo {author} {\bibfnamefont {M.}~\bibnamefont {M\aa{}nsson}}, \bibinfo
  {author} {\bibfnamefont {A.~L.}\ \bibnamefont {Chernyshev}}, \bibinfo
  {author} {\bibfnamefont {A.}~\bibnamefont {Mohan}}, \bibinfo {author}
  {\bibfnamefont {S.}~\bibnamefont {Singh}}, \bibinfo {author} {\bibfnamefont
  {C.}~\bibnamefont {Hess}}, \bibinfo {author} {\bibfnamefont {A.~T.}\
  \bibnamefont {Savici}}, \bibinfo {author} {\bibfnamefont {A.~I.}\
  \bibnamefont {Kolesnikov}}, \bibinfo {author} {\bibfnamefont
  {A.}~\bibnamefont {Piovano}}, \bibinfo {author} {\bibfnamefont
  {T.}~\bibnamefont {Perring}}, \bibinfo {author} {\bibfnamefont
  {I.}~\bibnamefont {Zaliznyak}}, \bibinfo {author} {\bibfnamefont
  {B.}~\bibnamefont {B\"uchner}}, \ and\ \bibinfo {author} {\bibfnamefont
  {A.}~\bibnamefont {Zheludev}},\ }\href {\doibase
  10.1103/PhysRevLett.111.067204} {\bibfield  {journal} {\bibinfo  {journal}
  {Phys. Rev. Lett.}\ }\textbf {\bibinfo {volume} {111}},\ \bibinfo {pages}
  {067204} (\bibinfo {year} {2013})}\BibitemShut {NoStop}%
\bibitem [{\citenamefont {Karmakar}\ \emph {et~al.}(2015)\citenamefont
  {Karmakar}, \citenamefont {Skoulatos}, \citenamefont {R\"uegg},\ and\
  \citenamefont {Singh}}]{Karmakar38}%
  \BibitemOpen
  \bibfield  {author} {\bibinfo {author} {\bibfnamefont {K.}~\bibnamefont
  {Karmakar}}, \bibinfo {author} {\bibfnamefont {M.}~\bibnamefont {Skoulatos}},
  \bibinfo {author} {\bibfnamefont {C.}~\bibnamefont {R\"uegg}}, \ and\
  \bibinfo {author} {\bibfnamefont {S.}~\bibnamefont {Singh}},\ }\href@noop {}
  {} (\bibinfo {year} {2015}),\ \bibinfo {note} {to be published}\BibitemShut
  {NoStop}%
\bibitem [{\citenamefont {Hammerath}\ \emph {et~al.}(2014)\citenamefont
  {Hammerath}, \citenamefont {Br\"uning}, \citenamefont {Sanna}, \citenamefont
  {Utz}, \citenamefont {Beesetty}, \citenamefont {Saint-Martin}, \citenamefont
  {Revcolevschi}, \citenamefont {Hess}, \citenamefont {B\"uchner},\ and\
  \citenamefont {Grafe}}]{Hammerath39}%
  \BibitemOpen
  \bibfield  {author} {\bibinfo {author} {\bibfnamefont {F.}~\bibnamefont
  {Hammerath}}, \bibinfo {author} {\bibfnamefont {E.~M.}\ \bibnamefont
  {Br\"uning}}, \bibinfo {author} {\bibfnamefont {S.}~\bibnamefont {Sanna}},
  \bibinfo {author} {\bibfnamefont {Y.}~\bibnamefont {Utz}}, \bibinfo {author}
  {\bibfnamefont {N.~S.}\ \bibnamefont {Beesetty}}, \bibinfo {author}
  {\bibfnamefont {R.}~\bibnamefont {Saint-Martin}}, \bibinfo {author}
  {\bibfnamefont {A.}~\bibnamefont {Revcolevschi}}, \bibinfo {author}
  {\bibfnamefont {C.}~\bibnamefont {Hess}}, \bibinfo {author} {\bibfnamefont
  {B.}~\bibnamefont {B\"uchner}}, \ and\ \bibinfo {author} {\bibfnamefont
  {H.-J.}\ \bibnamefont {Grafe}},\ }\href {\doibase 10.1103/PhysRevB.89.184410}
  {\bibfield  {journal} {\bibinfo  {journal} {Phys. Rev. B}\ }\textbf {\bibinfo
  {volume} {89}},\ \bibinfo {pages} {184410} (\bibinfo {year}
  {2014})}\BibitemShut {NoStop}%
\bibitem [{\citenamefont {Hammerath}\ \emph {et~al.}(2011)\citenamefont
  {Hammerath}, \citenamefont {Nishimoto}, \citenamefont {Grafe}, \citenamefont
  {Wolter}, \citenamefont {Kataev}, \citenamefont {Ribeiro}, \citenamefont
  {Hess}, \citenamefont {Drechsler},\ and\ \citenamefont
  {B\"uchner}}]{Hammerath40}%
  \BibitemOpen
  \bibfield  {author} {\bibinfo {author} {\bibfnamefont {F.}~\bibnamefont
  {Hammerath}}, \bibinfo {author} {\bibfnamefont {S.}~\bibnamefont
  {Nishimoto}}, \bibinfo {author} {\bibfnamefont {H.-J.}\ \bibnamefont
  {Grafe}}, \bibinfo {author} {\bibfnamefont {A.~U.~B.}\ \bibnamefont
  {Wolter}}, \bibinfo {author} {\bibfnamefont {V.}~\bibnamefont {Kataev}},
  \bibinfo {author} {\bibfnamefont {P.}~\bibnamefont {Ribeiro}}, \bibinfo
  {author} {\bibfnamefont {C.}~\bibnamefont {Hess}}, \bibinfo {author}
  {\bibfnamefont {S.-L.}\ \bibnamefont {Drechsler}}, \ and\ \bibinfo {author}
  {\bibfnamefont {B.}~\bibnamefont {B\"uchner}},\ }\href {\doibase
  10.1103/PhysRevLett.107.017203} {\bibfield  {journal} {\bibinfo  {journal}
  {Phys. Rev. Lett.}\ }\textbf {\bibinfo {volume} {107}},\ \bibinfo {pages}
  {017203} (\bibinfo {year} {2011})}\BibitemShut {NoStop}%
\bibitem [{\citenamefont {Eggert}\ \emph {et~al.}(2002)\citenamefont {Eggert},
  \citenamefont {Affleck},\ and\ \citenamefont {Horton}}]{Eggert29}%
  \BibitemOpen
  \bibfield  {author} {\bibinfo {author} {\bibfnamefont {S.}~\bibnamefont
  {Eggert}}, \bibinfo {author} {\bibfnamefont {I.}~\bibnamefont {Affleck}}, \
  and\ \bibinfo {author} {\bibfnamefont {M.~D.~P.}\ \bibnamefont {Horton}},\
  }\href {\doibase 10.1103/PhysRevLett.89.047202} {\bibfield  {journal}
  {\bibinfo  {journal} {Phys. Rev. Lett.}\ }\textbf {\bibinfo {volume} {89}},\
  \bibinfo {pages} {047202} (\bibinfo {year} {2002})}\BibitemShut {NoStop}%
\end{thebibliography}%
\end{document}